\documentclass[aps,prl,reprint,amsmath,amssymb,floatfix, footinbib]{revtex4-2}
\usepackage{amsthm,graphicx,tikz,pgfplots,needspace,bbm}
\usetikzlibrary{calc,arrows.meta}
\usepackage[mode=tex]{standalone}
\usepackage[colorlinks=true,citecolor=blue,linkcolor=blue,urlcolor=blue]{hyperref}
\hypersetup{pdftitle={Emergent Anyons in a Weakly Interacting Gauged Chern Insulator},pdfauthor={Sven Bachmann, Leonardo Goller, Marcello Porta}}
\pgfplotsset{compat=1.18}
\pgfplotsset{compat=1.18}
\usetikzlibrary{arrows.meta,decorations.markings,calc}
\definecolor{basepath}{RGB}{27,99,154}
\definecolor{liftpath}{RGB}{169,49,77}
\definecolor{fibcolour}{RGB}{123,106,147}
\pgfdeclarelayer{background layer}
\pgfdeclarelayer{foreground layer}
\pgfsetlayers{background layer,main,foreground layer}
\tikzset{
 ->-/.style={postaction={decorate},decoration={markings,
 mark=at position .5 with {\arrow{>}}}},
 ->>-/.style={postaction={decorate},decoration={markings,
 mark=at position .5 with {\arrow{>>}}}}
}
\renewcommand{\L}{\Lambda}
\DeclareMathOperator{\Tr}{Tr}
\newcommand{\dd}{\mathrm d}
\newtheorem{thm}{Theorem}
\newtheorem{lemma}[thm]{Lemma}
\newtheorem{proposition}[thm]{Proposition}

\newtheorem{defn}[thm]{Definition}
\theoremstyle{remark}\newtheorem{remark}[thm]{Remark}
\usepackage{comment}
\usepackage{physics, braket}
\begin{document}
\title{Emergent Anyons in a Weakly Interacting Gauged Chern Insulator}
\author{Sven Bachmann}
\affiliation{Department of Mathematics, The University of British Columbia, Vancouver, BC V6T 1Z2, Canada}
\author{Leonardo Goller}
\affiliation{%
Mathematics Area, SISSA -- International School for Advanced Studies,
Via Bonomea 265, 34136 Trieste, Italy
}
\begin{comment}\affiliation{%
IHES -- Institut des Hautes Études Scientifiques,
35 Route de Chartres, 91440 Bures-sur-Yvette, France
}
\end{comment}
\author{Marcello Porta}
\affiliation{%
Mathematics Area, SISSA -- International School for Advanced Studies,
Via Bonomea 265, 34136 Trieste, Italy
}
\date{\today}

\begin{abstract}
We derive a microscopic relation between the many-body Chern number $C$ of the fermionic
ground state and the braiding of fluxes in $\mathbb{Z}_{2}$ gauge theory coupled to matter: under a spectral gap condition, dressed flux loops obey $ W_2W_1=e^{i\frac{\pi}{2} C}W_1W_2$ on the ground-state
sector. We realize $C=1$ by gauging charge parity in a weakly interacting Qi–Wu–Zhang
insulator. Reflection positivity and chessboard estimates show that flux defects are massive, which yields a gapped, topologically ordered phase with four locally indistinguishable ground states.
\end{abstract}
\maketitle

\noindent \emph{Introduction.}
Topologically ordered phases provide one of the most striking
manifestations of collective quantum behavior
~\cite{Wen1990TopologicalOI}.
Unlike symmetry-breaking phases, they cannot be characterized
by a local order parameter and are instead detected through
robust ground-state degeneracy, local indistinguishability
~\cite{WenNiuTQO,kitaev2003}, and quasiparticle excitations
with nontrivial braiding statistics
~\cite{Leinaas:1977fm,wilczek1982magnetic,frohlich1976new}.
Exactly solvable lattice models have played a central role
in establishing this paradigm
~\cite{kitaev2003,kitaev2006,levin2005string},
but proving topological order and anyonic statistics in realistic 
interacting, non exactly solvable microscopic systems remains
considerably more difficult.
The main challenges are first to rigorously establish the presence of a spectral gap, and second to connect the statistics of excitations
to the geometry of the many-body ground-state space.

A natural route toward such systems is provided by gauged topological insulators.
Gauging a subgroup of the global $U(1)$ charge symmetry
promotes symmetry fluxes to physical excitations whose
statistics reflect the topological response of the fermions.
At the level of effective field theory, this mechanism is
captured by Chern--Simons theory 
~\cite{Froehlich, QHZ2008, ChengGu2014, WenNiuTQO,FroehlichZee,FroehlichStuder}. A microscopic derivation, however, requires control of the preferred gauge-flux sector, the energy cost
of flux defects, the persistence of a topological ground-state
manifold, and the operators implementing flux transport.

In~\cite{BGP}, using reflection positivity and quasi-adiabatic tools, we established a microscopic
relation between the algebra of dressed gauge-flux loops
and the Hall response of the fermionic matter, identifying
their commutator with a ground-state holonomy. The matter ground state has
uniform $\pi$ flux and vanishing Chern number:
the dressed flux loops commute, and the resulting anyon
algebra is that of the toric code~\cite{kitaev2003}.
For related microscopic results connecting Hall response
and flux statistics, see~\cite{KapustinSopenko}. 

In this Letter we derive the general flux-loop relation
and realize its nonzero-Chern case, in a microscopic setting.
We consider the Qi--Wu--Zhang (QWZ) model~\cite{QWZ}
with a weak on-site Hubbard interaction and gauge the
$\mathbb Z_2$-subgroup of its $U(1)$ charge symmetry. Reflection positivity selects a zero-flux
magnetic vacuum. Combined with the chessboard estimate, it allows us to quantify the energy cost of flux defects, yielding a positive flux mass. The fermionic gap remains open under weak interactions
~\cite{DeRoeckSalmhofer}, preserving the Chern number $C=1$
and the quantized Hall response~\cite{GMP}.
A uniform gap for all boundary twists then allows us to
construct quasi-adiabatic spectral flow, control the
splitting of the four physical ground states, and establish
their local indistinguishability.

The closest exactly solvable analogue is Kitaev's honeycomb
model~\cite{kitaev2006}. There too, fermions hop in a static
$\mathbb{Z}_2$ gauge field with conserved plaquette fluxes, and the
Chern number $\nu$ of the fermion bands fixes the anyon content through
the sixteenfold way. Our fermions are complex rather than Majorana, so
that $\nu=2C$: the gauged $C=1$ insulator realizes the Abelian $\nu=2$
entry of the sixteenfold way. The original
honeycomb model, with $\nu\in\{0,\pm1\}$, does not reach it; exactly
solvable generalizations do~\cite{Zhang16fold}, but only in
$\pi$-flux or vison-crystal sectors whose stability is established
numerically, since Lieb's theorem no longer applies to them. The two settings also differ in
what can be proved. In~\cite{kitaev2006} each flux sector is solved
exactly, but the vortex-free ground state is inferred from Lieb's
flux-phase theorem~\cite{LiebFlux}, and vortex energies are obtained
numerically or perturbatively; to our knowledge no uniform vortex mass,
and hence no uniform many-body gap, has been established. Here the
matter is interacting and time reversal is broken from the outset, yet
we prove that the zero-flux sector is the magnetic vacuum, give an
explicit lower bound on the energy of every flux defect, and combine it
with a uniform fermionic gap into a uniform physical gap.

The geometric content of the construction appears in the
algebra of dressed flux loops.
The underlying physical picture is that adiabatic flux
insertion, followed by a gauge transformation, can connect
distinct topological ground states
~\cite{ParamekantiVishwanath2004,OshikawaSenthil2006}. We implement this process using Hastings' quasi-adiabatic continuation~\cite{HW,BMNS},
which realizes parallel transport in the many-body
ground-state line bundle over the boundary-condition
torus~\cite{NTW,AvronSeiler1985}.
Gauge covariance identifies the commutator of two
noncontractible flux loops with the holonomy around
the boundary of a quarter of this torus.
Locality makes the Berry curvature asymptotically independent
of the twists, fixing the commutation phase to $e^{i \frac{\pi}{2}C}$,
where $C$ is the many-body Chern number.
For $C=1$, the gauged interacting Chern insulator thus realizes a nontrivial vortex braiding algebra starting from an explicit microscopic Hamiltonian. \\

\noindent \emph{The model.}
Let $\Lambda_L=(\mathbb Z/L\mathbb Z)^2$, $L\in4\mathbb N$.
Write $c^{\pm}_{x,\alpha}$ for the fermionic creation and annihilation operators, $\alpha=1,2$, and place one
gauge qubit on each nearest-neighbour bond. The bond operators $\widehat\sigma_{xy}^3$ represent the
$\mathbb Z_2$-gauge potential, while $\widehat\sigma_{xy}^1$
is the conjugate electric-field operator. In compact form,
\begin{equation}
 H_\lambda=\sum_{x,y,\alpha,\beta}
 c^+_{x,\alpha}h(\widehat\sigma)_{x\alpha,y\beta}c^-_{y,\beta}
 +\lambda \sum_{x} (n_{x, 1}-\frac{1}{2})(n_{x,2}-\frac{1}{2})\label{eq:model}
\end{equation}
with, for $t>0$ and $m\in\mathbb R$, 
\begin{equation}
h(\widehat\sigma)=m\mathbbm{1}\otimes\tau^3
 -\frac{t}{2}\sum_{\mu=1}^2
 \bigl[ T_\mu(\hat{\sigma})\otimes D_\mu+\mathrm{h.c.}\bigr],
 \label{eq:one-body-hamiltonian}
\end{equation}
where $\tau^\mu$ are Pauli matrices acting on the spin degrees of freedom of the fermions and
\begin{align*}
 ( T_\mu(\widehat{\sigma}))_{xy}
 &= \widehat\sigma^3_{x,x+e_\mu}\delta_{y,x+e_\mu}
 \quad D_{\mu} = \tau^3 - i \tau^{\mu}.
\end{align*}
Physical states satisfy the local $\mathbb Z_2$-Gauss law,
\begin{equation}
 G_x|\Psi\rangle=|\Psi\rangle,
 \qquad G_x=(-1)^{N_x} A_x,
 \label{eq:gauss}
\end{equation}
where $N_x=\sum_\alpha n_{x,\alpha}$ and $A_x = \prod_{\langle x,y\rangle\ni x} \widehat\sigma_{xy}^1$.
In the absence of an electric-field term, the magnetic
fluxes
$B_p=\prod_{\langle x,y\rangle\in\partial p}
\widehat\sigma_{xy}^3$
are conserved.
A plaquette with $B_p=+1$ carries zero flux, whereas
$B_p=-1$ corresponds to a magnetic vortex, or vison.
We can therefore study the Hamiltonians
$H_\lambda(\sigma)$ in fixed bond backgrounds
$\sigma_{xy}=\pm1$.
Gauge-equivalent backgrounds yield unitarily equivalent
Hamiltonians: their spectra depend only on the plaquette
fluxes and the two noncontractible $\mathbb Z_2$-holonomies
on the torus, see~\cite{LiebLoss,GP}.
The physical spectrum in each gauge class is obtained by
restricting $H_\lambda(\sigma)$ to even fermion parity %
\footnote{Let $P_G=\prod_x(1+G_x)/2$ and $N=\sum_xN_x$.
For a normalized bond state $|\sigma\rangle$ and any
fermionic state $|\psi\rangle$, connectedness of the lattice gives
\[
 \bigl\|P_G(|\psi\rangle\otimes|\sigma\rangle)\bigr\|^2
 =2^{-|\Lambda_L|}
 \langle\psi|[1+(-1)^N]|\psi\rangle.
\]
Thus odd-parity states are annihilated, while for even
$|\psi\rangle$ the squared norm is
$2^{1-|\Lambda_L|}\langle\psi|\psi\rangle$.
The normalized map
$|\psi\rangle\mapsto
2^{(|\Lambda_L|-1)/2}
P_G(|\psi\rangle\otimes|\sigma\rangle)$
is a unitary identification with the physical sector
of the gauge orbit (for details see Proposition~2.7 and
Remark~2.8 of~\cite{GP}).}. Finally, the torus identity $\prod_pB_p=\mathbbm{1}$
implies that magnetic vortices occur in even numbers.

\begin{figure}
\centering
\begin{tikzpicture}[x=1cm, y=1cm, font=\footnotesize]
  \definecolor{flux}{HTML}{C8742B}   % flux-defect sectors
  \definecolor{ferm}{HTML}{3A6EA5}   % fermionic excitations
  \definecolor{gs}{HTML}{1F1F1F}     % ground sectors
  % geometry: band columns, thresholds (lower bounds), band tops
  \def\xL{0.45}  \def\xM{2.20}  \def\xR{3.95}
  \def\yV{1.30}  \def\yF{2.30}  \def\yT{3.75}
  % ------------------------------------------------------------ axis + ticks
  \draw[-{Stealth[length=2mm]}, thick] (0,-0.12) -- (0,\yT+0.75)
        node[above] {$E-E_{\mathrm{ref},\lambda}$};
  \foreach \y/\lab in {0/$0$, \yV/$2d_{*,\lambda}$, \yF/$M_\infty(\lambda)$}
    { \draw (0,\y) -- (-0.08,\y) node[left] {\lab}; }
  % threshold guides (drawn first, so the bands cover them)
  \draw[ferm!80!black, densely dashed, thin] (0,\yF) -- (\xR,\yF);
  \draw[flux!90!black, densely dashed, thin] (0,\yV) -- (\xM-0.06,\yV);
  % ------------------------------------------------------------ flux-defect band
  \shade[bottom color=flux!32, top color=flux!3]
        (\xL,\yV+0.05) rectangle (\xM-0.06,\yT);
  \foreach \y/\o in {0.090/0.97,0.256/0.92,0.477/0.85,0.676/0.78,0.818/0.74,0.970/0.69,1.107/0.64,1.248/0.60,1.387/0.55,1.478/0.52,1.558/0.50,1.673/0.46,1.760/0.43,1.856/0.40,1.916/0.38,1.987/0.36,2.063/0.34,2.117/0.32,2.190/0.30,2.256/0.27,2.295/0.26,2.331/0.25}
    { \draw[flux!75!black, line width=0.3pt, opacity=\o]
        (\xL,\yV+\y) -- (\xM-0.06,\yV+\y); }
  % ------------------------------------------------------------ fermionic band
  \shade[bottom color=ferm!30, top color=ferm!3]
        (\xM+0.06,\yF+0.05) rectangle (\xR,\yT);
  \foreach \y/\o in {0.085/0.95,0.249/0.86,0.399/0.77,0.487/0.72,0.570/0.68,0.682/0.61,0.780/0.56,0.868/0.50,0.935/0.47,1.008/0.42,1.075/0.39,1.135/0.35,1.180/0.33,1.226/0.30,1.269/0.28,1.315/0.25}
    { \draw[ferm!75!black, line width=0.3pt, opacity=\o]
        (\xM+0.06,\yF+\y) -- (\xR,\yF+\y); }
  % band labels, above the bands
  \node[align=center, text=flux!65!black, anchor=south]
        at ($(\xL,\yT)!0.5!(\xM-0.06,\yT)$) {flux-defect\\[-1pt]sectors};
  \node[align=center, text=ferm!65!black, anchor=south]
        at ($(\xM+0.06,\yT)!0.5!(\xR,\yT)$) {fermionic\\[-1pt]excitations};
  % ------------------------------------------------------------ ground sectors
  \foreach \y in {0, 0.018, 0.031, 0.052}
    { \draw[gs, line width=0.45pt] (\xL,\y) -- (\xR,\y); }
  \node[below, align=center] at ($(\xL,0)!0.5!(\xR,0)+(0.12,-0.06)$)
        {ground sectors $|\Omega_{a,b}\rangle$, $a,b=\pm1$};
  % ------------------------------------------------------------ gap
  \draw[{Stealth[length=1.7mm]}-{Stealth[length=1.7mm]}, thick]
        (1.33,0.10) -- (1.33,\yV-0.03)
        node[midway, right=1pt] {$g_{\mathrm{phys}}$};
  % ------------------------------------------------------------ magnifier
  \draw[black!55, thin] (3.95,0.026) circle (0.12);
  \draw[black!35, thin] (3.876,0.120) -- (5.581,1.455);
  \draw[black!35, thin] (3.969,-0.093) -- (6.107,0.248);
  \fill[white] (6.0,0.92) circle (0.68);
  \begin{scope}
    \clip (6.0,0.92) circle (0.68);
    \foreach \y in {0, 0.122, 0.211, 0.354}
      { \draw[gs, line width=0.6pt] (5.2,0.743+\y) -- (6.10,0.743+\y); }
  \end{scope}
  \draw[{Stealth[length=1.3mm]}-{Stealth[length=1.3mm]}, thin]
        (6.20,0.743) -- (6.20,1.097) node[midway, right=-1pt] {$\epsilon_L$};
  \draw[black!55, thin] (6.0,0.92) circle (0.68);
  \node at (6.0,1.83) {$\epsilon_L=O(L^{-\infty})$};
\end{tikzpicture}
 \caption{Schematic energy spectrum in the gapped phase.
 The four low-energy states $|\Omega_{a,b}\rangle$,
 labelled by the holonomies $a,b=\pm1$, have energy spread
 $\epsilon_L=O(L^{-\infty})$.
 They are separated from the rest of the spectrum by a gap $g_{\mathrm{phys}}$.
 }
 \label{fig:physical-spectrum}
\end{figure}
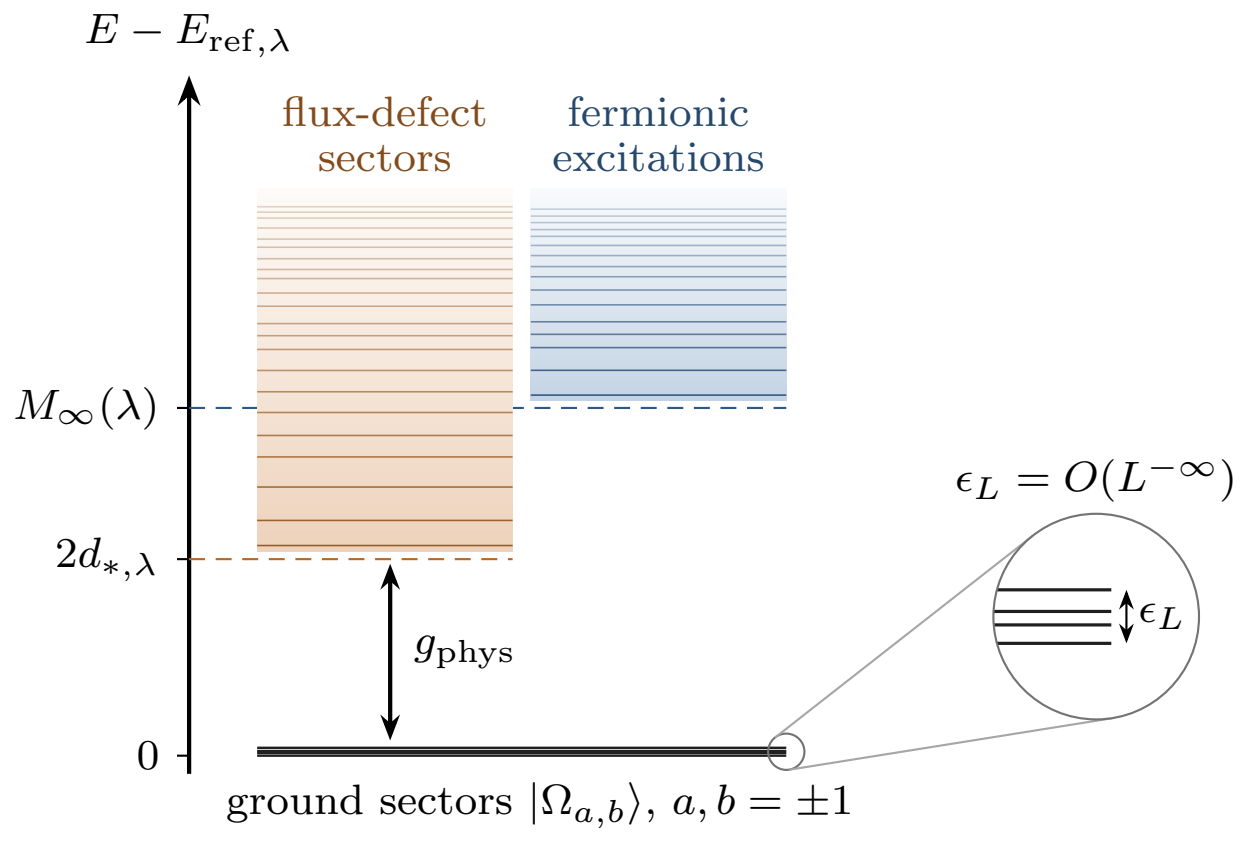

Our analysis relies on a generalization of the reflection-positivity
strategy of Lieb's flux-phase theorem~\cite{LiebFlux}
and its $\mathbb{Z}_{2}$ gauge-theory implementations~\cite{GP,BGP}. A main difference with respect to these works is that, for the QWZ model, the minimum of the
half-filled matter ground-state energy over bond backgrounds is attained in a flat configuration $B_p=+1$, with antiperiodic
boundary conditions in both directions.
We first establish the fermionic gap and topology
of the flat sectors under weak interactions.
Chessboard estimates then reduce the comparison with nonflat backgrounds to a periodic-cell calculation, yielding a positive energy cost for the vison excitations. Together, these estimates isolate the four physical ground states, and show that charge and magnetic excitations are massive, see Figure~\ref{fig:physical-spectrum}. \\

\noindent \emph{Gap and topology of the magnetic vacuum.}
The four flat gauge classes are labeled by two holonomies $a,b=\pm1$, corresponding to periodic
or antiperiodic boundary conditions for the fermions. Allowed momenta $k = (k_{1}, k_{2})$ in $\Lambda_{L,(a,b)}^*$ satisfy $e^{iLk_1}=a$ and $e^{iLk_2}=b$. At $\lambda=0$, the Bloch Hamiltonian reads
\begin{equation}
    h(k)=\vec v(k) \cdot \vec \tau,
\end{equation}
where 
\begin{equation}
    \vec v(k) = (-t\sin k_1,-t\sin k_2, m -t(\cos k_1+\cos k_2)).
\end{equation}
Since $h(k)^2 = \vert \vec v(k)\vert^2 \mathbbm{1}$, the eigenvalues are $\pm \vert \vec v(k)\vert$, namely 
\begin{equation}
 E(k)^2
 =t^2(\sin^2 k_1+\sin^2 k_2)
       +[m-t(\cos k_1+\cos k_2)]^2.
       \label{eq:disp_rel}
\end{equation}
The gap between the energy bands is 
\begin{equation}
 M_\infty(0)=\min_{k\in\mathbb T^2}E(k)
 =\min\{|m-2t|,|m|,|m+2t|\}.
 \label{eq:gap-free}
\end{equation}
For $M_\infty(0)>0$, the free ground state fills
the lower band and has $N=L^2$ particles. Since the Gauss projection restricts the matter states
to even fermion parity, the gap within the physical
flat sector is bounded below by $2M_\infty(0)$ and converges
to this value in the thermodynamic limit.

\begin{figure}
\centering
\begin{tikzpicture}[x=0.82cm, y=0.82cm, font=\footnotesize]
  \definecolor{halfplus}{HTML}{3A6EA5}   % blue of Figs. 1, 3
  \definecolor{plane}{HTML}{C8742B}      % brown of Figs. 1, 3
  \def\M{3.2}  \def\T{2.7}               % plotted window: |m| <= M, 0 <= t <= T
  % ------------------------------------------------------------ phases
  \fill[halfplus!24] (0,0) -- (-\M,\M/2) -- (-\M,\T) -- (0,\T) -- cycle;
  \fill[plane!30]    (0,0) -- ( \M,\M/2) -- ( \M,\T) -- (0,\T) -- cycle;
  % ------------------------------------------------------------ axes
  \draw[-{Stealth[length=1.8mm]}, line width=0.5pt] (-\M-0.25,0) -- (\M+0.45,0)
        node[right] {$m$};
  \draw[-{Stealth[length=1.8mm]}, line width=0.5pt] (0,0) -- (0,\T+0.45)
        node[above] {$t$};
  \node[below] at (0,0) {$0$};
  % ------------------------------------------------------------ gapless rays
  \draw[line width=0.9pt, line cap=round] (0,0) -- (-\M,\M/2)
        node[left=1pt] {$m=-2t$};
  \draw[line width=0.9pt, line cap=round] (0,0) -- ( \M,\M/2)
        node[right=1pt] {$m=2t$};
  \draw[line width=0.9pt, line cap=round] (0,0) -- (0,\T);
  \node[fill=white, inner sep=1.5pt, rounded corners=1pt] at (0,1.05) {$m=0$};
  % ------------------------------------------------------------ labels
  \node[text=halfplus!75!black] at (-1.6,2.15) {$C=+1$};
  \node[text=plane!70!black]    at ( 1.6,2.15) {$C=-1$};
  \node at (-2.25,0.45) {$C=0$};
  \node at ( 2.25,0.45) {$C=0$};
\end{tikzpicture}
\caption{Free phase diagram in the zero-flux sector.
 The three boundary rays are gapless.
 Weak interactions preserve the Chern number on compact
 subsets of the open phases; the interacting transition
 curves are not determined here.}
    \label{fig:phase}
\end{figure}
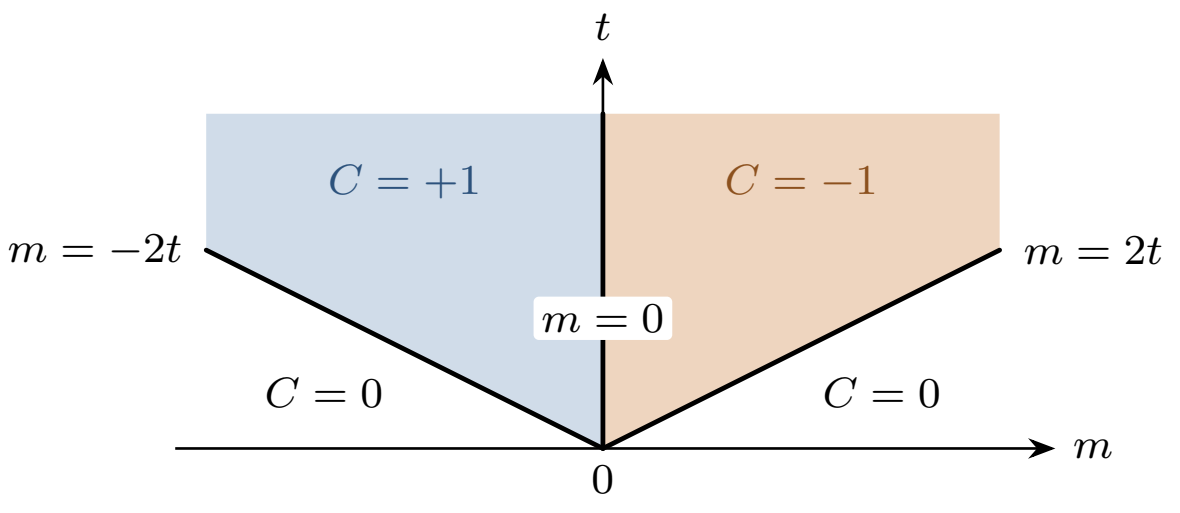

In a gapped phase, the one-body Berry curvature is given by
\begin{equation}\label{1-p curvature}
    \Omega(k)=\frac12\,\hat v(k)\cdot\bigl(\partial_{1}\hat v(k)\times\partial_{2}\hat v(k)\bigr),
\quad \hat v=\vec v/|\vec v|,
\end{equation}
and the Chern number of the occupied band is:
\begin{equation}\label{eq:chern-free}
\begin{split}
    \kappa&=\frac{1}{2\pi}\int_{\mathbb T^2}\Omega(k)\,d^2k \\
      &= \frac12\bigl[\operatorname{sgn}(m-2t)-2\operatorname{sgn}m+\operatorname{sgn}(m+2t)\bigr],
      \end{split}
\end{equation}
which yields the free phase diagram of Figure~\ref{fig:phase}.

In order to describe flux insertion, we extend the matter
Hamiltonian to continuous boundary twists
$\varphi\in\mathbb T^2$.
These insert flux through the noncontractible cycles
while leaving the plaquette fluxes unchanged.
In the free model, these fluxes can be introduced via a shift of the momentum grid
by $\varphi_j/L$, preserving the lower bound
$M_\infty(0)$ for the Fock-space gap.

By~\cite{DeRoeckSalmhofer}, the gap remains open for sufficiently small $|\lambda|$. In fact, the gap is bounded below by a
twist-independent, volume independent constant $ M_\infty(\lambda)
 \geq M_\infty(0)-c|\lambda|^{1/3}>0$. At periodic or antiperiodic boundary conditions, particle-number conservation, uniqueness of the ground state, and
particle--hole symmetry imply $N=L^2$ \footnote{
The unitary particle--hole transformation
$\mathcal U_{\mathrm{ph}}$ acts as
\[
 \mathcal U_{\mathrm{ph}}c^\pm_{x,\alpha}
 \mathcal U_{\mathrm{ph}}^{-1}
 =\sum_\beta(\tau^1)_{\alpha\beta}c^\mp_{x,\beta},
\]
and maps $N$ to $2L^2-N$.
For real bond backgrounds it leaves $H_\lambda(\sigma)$
invariant, including the centered Hubbard interaction.
For continuous twists it relates
$H_\lambda(\varphi)$ to $H_\lambda(-\varphi)$.}. Since the ground state remains unique and gapped
across the boundary-condition torus, its particle
number cannot change as the twists vary.
Consequently, the ground state is half-filled
for every $\varphi\in\mathbb T^2$.
For $L\in4\mathbb N$, all these states have even
fermion parity and are therefore admissible as physical states.

The uniform gap allows to define a fermionic ground-state line bundle over the torus of boundary twists. For the rank-one matter ground-state projector
$P (\varphi)$ in a reference flat sector, the Berry curvature 
\begin{equation}
    F_\lambda(\varphi)=i\operatorname{Tr}\bigl(P(\varphi)[\partial_1P(\varphi),\partial_2P(\varphi)]\bigr),
\end{equation}
yields the many-body Chern number~\cite{NTW,AvronSeiler1985}
\begin{equation}
C_\lambda=\frac{1}{2\pi}\int_{\mathbb T^2}F_\lambda(\varphi)\,d^2\varphi,
 \label{eq:chern-interacting}
\end{equation}
where the trace is over the fermionic Fock space. Changing the flat-sector holonomies translates
the twists by $\pi$ in the corresponding directions,
leaving the integral over the full twist torus unchanged. At $\lambda=0$, the half-filled ground state is
a Slater determinant, with:
\begin{equation}\label{eq:F0phi}
F_0(\varphi)=\frac1{L^2}\sum_{n_1,n_2=0}^{L-1}
       \Omega\left(\frac{2\pi n+\varphi}{L}\right).
\end{equation}
Integrating (\ref{eq:F0phi}) over the twist torus one gets $C_0=\kappa \in \mathbb{Z}$. As long as the gap stays open, the projector is smooth in the twists and $\lambda$, and hence its first Chern number is a constant along this interpolation, namely $C_\lambda = \kappa$. The uniform gap and locality further yield constant curvature, see ~\cite{GMP} and~\cite{HastingsMichalakis,BBDF}:
\begin{equation}
 \sup_\varphi|F_\lambda(\varphi)-\kappa/(2\pi)| = O(L^{-\infty}).
 \label{sm:flat-curvature}
\end{equation}

The matter ground states $\ket{\psi_{a,b}}$ in a flat gauge configuration $\sigma$ determine the physical
ground states $\ket{\Omega_{a,b}}$:
\begin{equation}
 \ket{\Omega_{a,b}}
 =2^{(L^2-1)/2}P_G
\bigl(\ket{\psi_{a,b}}\otimes\ket{\sigma}\bigr),
 \label{eq:physical-ground-states}
\end{equation}
where $P_G=\prod_x(1+G_x)/2$ is the Gauss law projection. It preserves the Berry connection of matter states, being a twist-independent isometry.

The four states $\ket{\psi_{a,b}}$ have exponentially close energies. For $|\lambda|$ small, locality and the uniform many-body gap give~\cite{PhysRevB.98.155137, BGP}, see also \cite{SM}:
\begin{equation}\label{eq:GS splitting}
     |E_{\rm{ref},\lambda}(\varphi)-E_{\rm{ref},\lambda}(\varphi')|
 \leq O(L^{-\infty}),
 \quad (\varphi,\varphi'\in\mathbb T^2),
\end{equation}
where $E_{\rm{ref},\lambda}(\varphi)$ is the matter ground-state energy with boundary twist $\varphi$ upon a reference flat background. In particular, this shows that the four inequivalent ground states are spread by $\epsilon_L=O(L^{-\infty})$.\\

\noindent \emph{Reflection positivity and the chessboard bound.}
To prove that the ground state vectors lie in the flat sector and to control the energy cost of magnetic defects, we shall rely on reflection positivity for the matter Hamiltonians
$H_\lambda(\sigma)$, via an extension of the approach of~\cite{LiebFlux, GP, BGP}. The matter partition function is $Z_\lambda(\sigma)=\Tr e^{-\beta H_\lambda(\sigma)}$ and the ground state energy is $-\lim_{\beta \to \infty}\beta^{-1}\log Z_\lambda(\sigma)$.

\begin{figure}
\centering
\input{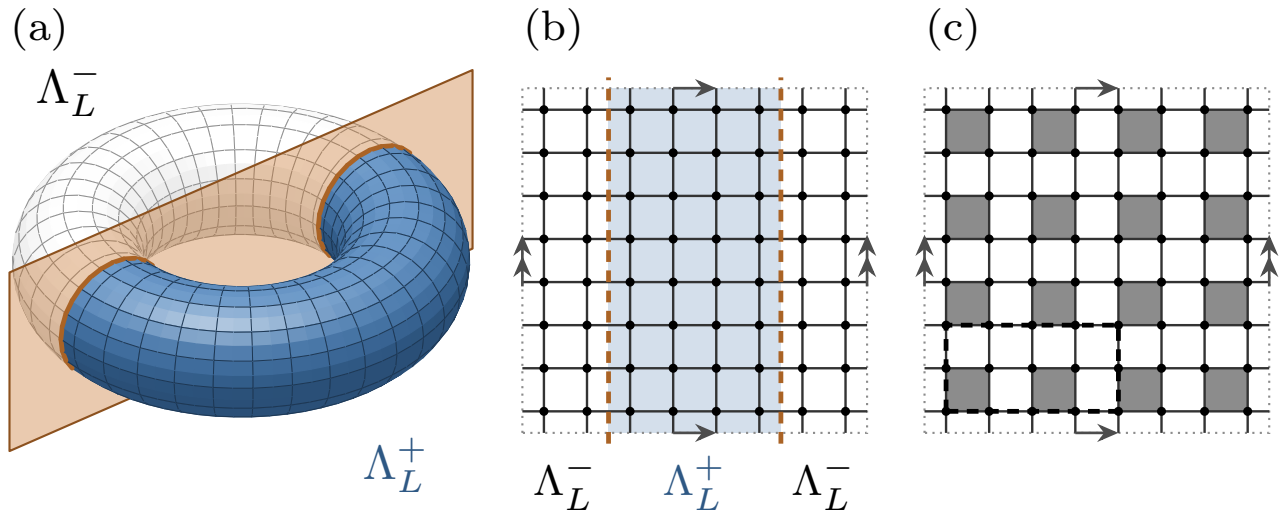}
\caption{Reflection geometry on the torus (a), its
periodic lattice representation (b), and the chessboard flux configuration (c).
The reflection exchanges $\Lambda_L^+$ and $\Lambda_L^-$. Shaded plaquettes carry a $\pi$-flux.}
\label{fig:cut1}
\end{figure}

We cut the physical torus in two halves $\Lambda_L^\pm$ mapped onto each other by a reflection $\vartheta$ across a plane, see Figure~\ref{fig:cut1}. Let $\mathcal A_\pm$ be the CAR algebras generated
by the fermionic fields supported on
$\Lambda_L^\pm$, respectively. The reflection $\Theta$ is a
parity-preserving, antilinear $*$-automorphism of $\mathcal A$ such that $\Theta^2=\mathrm{id}$ and $\Theta(\mathcal A_\pm)=\mathcal A_\mp$. For homogeneous $A,B\in\mathcal A_+$, define
$\Theta(A)\circ B
=i^{p(A)p(B)}\Theta(A)B$, where $p(A)=0,1$ is the fermion parity of $A$,
extended by linearity.
This twisted product accounts for the graded
commutation of operators on opposite halves.

Suppose that the Hamiltonian $H$ has the form:
\begin{equation}
    H=H_++\Theta(H_-) -\sum_j\Theta(C_j)\circ D_j
\end{equation}
where $H_\pm\in\mathcal A_+$ are even and self-adjoint and $C_j,D_j\in\mathcal A_+$ are odd. Then with $H^{(-)}=H_-+\Theta(H_-)-\sum_j\Theta(C_j)\circ C_j$ and $H^{(+)}=H_++\Theta(H_+) -\sum_j\Theta(D_j)\circ D_j$, \emph{reflection positivity} means that for every $\beta>0$,
\begin{equation}
 \bigl(\Tr e^{-\beta H}\bigr)^2
 \leq
 \bigl(\Tr e^{-\beta H^{(-)}}\bigr)
 \bigl(\Tr e^{-\beta H^{(+)}}\bigr).
 \label{eq:reflection-bound}
\end{equation}

For the QWZ Hamiltonian, a bond reflection
$\vartheta_\nu$ perpendicular to an edge $e_\nu\:(\nu = 1,2)$ is implemented by
\begin{equation}
 \Theta_\nu(c^+_{x,\alpha})
 =\sum_\beta(\tau^\nu)_{\alpha\beta}
   c^-_{\vartheta_\nu(x),\beta}.
 \label{eq:reflection}
\end{equation}
Observe that, compared to \cite{LiebFlux,GP,BGP}, the geometric reflection and particle-hole transformation have to be combined with a transformation acting on the internal degrees of freedom. Then, for an arbitrary bond background, we choose a gauge
in which the bonds normal to the left boundary of
$\Lambda_L^+$ have sign $-1$, whereas those normal
to its right boundary have sign $+1$. The boundary coupling matrices are
$\mathbbm{1}\pm S_\nu$, with $S_\nu=-i\tau^\nu\tau^3$;
their eigenvalues are $0$ and $2$.
Diagonalization gives the boundary
decomposition required for reflection positivity \cite{SM}.
The on-site Hubbard term contributes only to the
half-Hamiltonians and preserves this structure
for every real $\lambda$. In this setting, \eqref{eq:reflection-bound} becomes
\begin{equation}\label{eq:RP}
Z_\lambda(\sigma)^2\leq Z_\lambda(\sigma^-)Z_\lambda(\sigma^+)    
\end{equation}
where $\sigma^\pm$ are obtained by retaining
and reflecting either half-background.
In each reflected configuration, the
normal bond signs occur in identical pairs, except at the seams where they remain unchanged. The holonomy perpendicular to the seams after reflection is therefore equal to $-1$: reflections in both directions yield antiperiodic
boundary conditions. Moreover, $B_p=+1$ for each plaquette cut by the seam.
This replaces the $\pi$-flux seams of \cite{LiebFlux, GP, BGP} with zero-flux seams. 

Repeated reflections disseminate a chosen defect while inserting rows and columns of zero flux resulting in the chessboard pattern~\footnote{For $L\in4\mathbb N$, this pattern is preserved by reflections across its zero-flux seams; thus, it carries antiperiodic boundary conditions in both directions.} of Figure~\ref{fig:cut1}. This observation and repeated applications of~\eqref{eq:RP} yield a lower bound for the ground state energy $E_{\lambda}(\sigma)$ of $H_\lambda(\sigma)$ for any background $\sigma$:
\begin{align}
 E_\lambda(\sigma)
 &\geq E_{{\rm ref},\lambda}
       +n_v(\sigma)d_{*,\lambda},\label{eq:chess0}\\
 d_{*,\lambda}
 &=\frac{E_{*,\lambda}
        -E_{{\rm ref},\lambda}}{L^2}.
 \label{eq:mass-criterion}
\end{align}
Here, $n_v(\sigma)$ is the number of vortices in the configuration $\sigma$, $E_{{\rm ref},\lambda}$ and  $E_{*,\lambda}$ are the ground state energies for the flat pattern and the chessboard pattern, with antiperiodic boundary conditions.

At $\lambda=0$, the periodicity of the chessboard configuration, with a $4\times2$ unit cell, allows to compute explicitly $d_{*,0}$ for large $L$~\cite{SM}. At $m=-t$, the numerical value is
$a_{0}t = 0.066875\,t$. Since the norm of the interaction is bounded above by $|\lambda| L^2/4$, the interaction changes
each ground-state energy density by at most
$|\lambda|/4$.
Together with the dependence on $m$, this gives
\begin{equation}
 \liminf_{L\to\infty}d_{*,\lambda}
 \geq a_0t-2|m+t|-\frac{|\lambda|}{2}.
 \label{eq:mass-stability}
\end{equation}
Hence flux defects have a positive energy cost in a neighbourhood of $m=-t,\lambda=0$. \\

\noindent \emph{Quasi-adiabatic flow and the physical ground space.} We now pick a flat background and denote $H_{\lambda}(\varphi)$ the Hamiltonian in that background with boundary twists $\varphi\in\mathbb T^2$. For $j=1,2$, the quasi-adiabatic generators $K_j(\varphi)$~\cite{SM} are such that $\partial_jP=i[K_j,P]$ and $PK_jP=0$ \cite{HW,BMNS}. Integrating $\dot V_\gamma(s)=i\dot{\vec \gamma}(s)\cdot \vec K(\gamma(s))V_\gamma(s)$ along a path $\gamma$ in flux space  with $V_\gamma(0)=\mathbbm{1}$ yields a unitary quasi-adiabatic flow which realizes the Berry parallel transport
\begin{equation}\label{eq:qa-flow}
    V_\gamma(s) P(0) = P(\gamma(s))V_\gamma(s).
\end{equation}
It preserves the Gauss law, particle number, and the magnetic background.

The twist derivative $\partial_jH_{\lambda}$ is supported along the boundary. A Lieb--Robinson bound and the rapidly decay of the function $f_g$ make $K_j$ quasi-local near that
cut. This locality depends crucially on the uniform gap in the flat sector, but not on the magnetic vortex gap.

The quasi-locality of $K_j$ has two important consequences \cite{SM}. Firstly, it provides the ground state energy splitting estimates~\eqref{eq:GS splitting}, as the following argument shows. Move the twist to a parallel
cut at distance $L/2$ from the current $\partial_j H_\lambda$. Changing the
twist then changes that current by $O(L^{-\infty})$. Periodicity of
the ground state energy forces its mean derivative to vanish, hence the
energy variation itself is $O(L^{-\infty})$. Secondly, the possibility of moving both cuts away from a local observable yields local indistinguishability. Consequently the four Gauss-projected flat ground states $|\Omega_{a,b}\rangle$ have energy spread $\epsilon_L=O(L^{-\infty})$ and they are isolated by a spectral gap (taking into account the even number of vortices)
\begin{equation}
 g_{{\rm phys}}\geq \min\{M_\infty(\lambda),2d_{*,\lambda}\}-\epsilon_L.
 \label{eq:physical-gap}
\end{equation}\\

\begin{figure}
\centering
\input{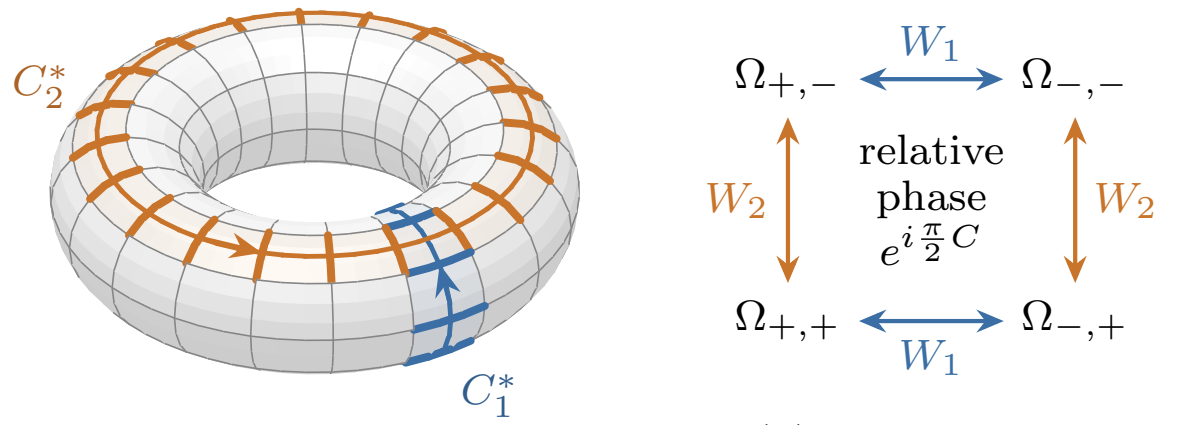}
\caption{(a) Two closed cuts on the lattice.
The colored bonds support the bare gauge flips $\widehat X_j$;
$V_{\alpha_j}$ dresses them in a quasi-local ribbon. (b) Action of the
dressed loops on the four physical ground states. The two routes from $\Omega_{+,+}$ to $\Omega_{-,-}$ differ by the
braiding phase.}
\label{fig:loops}
\end{figure}

\noindent \emph{Flux operators and transport.}
A closed dual string $\widehat X_j=\prod_{b\pitchfork C_j^*}
\widehat\sigma_b^1$ flips one holonomy, without creating plaquette
flux. Let us consider the operator $V_{\gamma_{j}}$ that implements the parallel transport~\eqref{eq:qa-flow} along the path $\gamma_j(s)=\pi se_j$, and define the operators:
\begin{equation}
 W_j=\widehat X_jV_{\gamma_j}.\label{eq:loops}
\end{equation}
They connect the four different ground states of the gauge theory:
$W_1|\Omega_{a,b}\rangle=c_1(a,b)|\Omega_{-a,b}\rangle$ and
$W_2|\Omega_{a,b}\rangle=c_2(a,b)|\Omega_{a,-b}\rangle$, where
$|c_j|=1$, see Fig.~\ref{fig:loops}.

\begin{figure}
\centering
\input{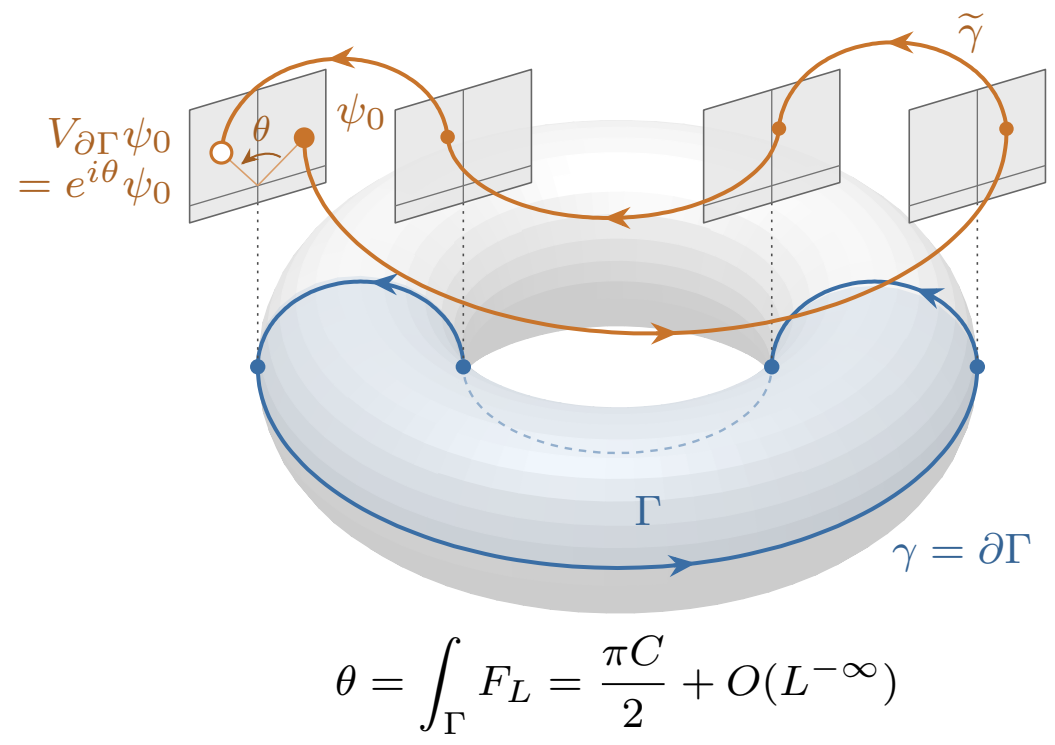}
 \caption{Ground-state line bundle over the boundary-condition
 torus. One-dimensional complex fibres are depicted as real planes
 in local charts. The blue loop $\gamma=\partial\Gamma$ has a
 horizontal lift $\widetilde\gamma$ under quasi-adiabatic transport, shown in brown.
 The transported vector returns to the initial fibre as
 $e^{i\theta}\psi_0$, with
 $\theta=\pi C/2+O(L^{-\infty})$.}
 \label{fig:bundle}
\end{figure}

Consider the ground-state line bundle over the twist torus $\mathbb T^2_\varphi$ (Fig.~\ref{fig:bundle}). A vector can
be chosen smoothly on a square, but for $C\ne0$ there
is no global smooth periodic ground-state vector. The curvature measures
this obstruction. Gauge covariance gives~\cite{SM}
\begin{equation}
 W_2^*W_1^*W_2W_1=V_{\partial\Gamma},
 \qquad \Gamma=[0,\pi]^2.\label{eq:quarter}
\end{equation}
Products act from right to left: the path visits $(0,0)$,
$(\pi,0)$, $(\pi,\pi)$, $(0,\pi)$, and returns to $(0,0)$.
On a ground state, its holonomy is
$\exp(i\int_\Gamma F_\lambda d\varphi)$. The asymptotically flat curvature~\eqref{sm:flat-curvature} yields
\begin{equation}\label{eq:braid}
  W_2^*W_1^*W_2W_1\Pi =e^{i \frac{\pi}{2}C}\Pi +O(L^{-\infty}),
\end{equation}
where $\Pi$ is the projector on the ground state sector.

%To conclude, we note that the loop commutator is a full monodromy. It should be distinguished from a single exchange: selecting a vortex charge branch gives the conventional exchange phase $e^{i\pi C/4}$~\cite{KapustinSopenko}, but the torus calculation alone does not construct a local exchange. \\

\noindent \emph{Use of AI Tools.}
ChatGPT 5.6, 6 and Claude Opus 5.5 were used as research assistants. ChatGPT pointed out the QWZ model as a nearest-neighbour lattice model with a nontrivial topological phase and it performed the numerical calculation of the magnetic gap. ChatGPT and Claude have been used to produce the source code of the figures, for proofreading and for a literature search. The proof strategy has been developed by the authors, without the use of AI tools. The authors take full responsibility for the scientific content and conclusions of this work.\\

\noindent \emph{Acknowledgements.} SB is supported by NSERC of Canada under Grant RGPIN-2024-03927. The work of MP has been carried out under the auspices of GNFM of INdAM.

\bibliography{QWZ_references}

\bigskip
\bigskip

\onecolumngrid
\setlength{\abovedisplayskip}{7pt plus 2pt minus 2pt}
\setlength{\belowdisplayskip}{7pt plus 2pt minus 2pt}
\setlength{\abovedisplayshortskip}{3pt plus 1pt}
\setlength{\belowdisplayshortskip}{5pt plus 1pt minus 1pt}
\setcounter{secnumdepth}{2}
\setcounter{section}{0}
\setcounter{subsection}{0}
\setcounter{equation}{0}
\setcounter{figure}{0}
\setcounter{thm}{0}
\renewcommand{\thesection}{S\arabic{section}}
\renewcommand{\thesubsection}{\Alph{subsection}}
\renewcommand{\theHsection}{S\arabic{section}}
\renewcommand{\theHsubsection}{S\arabic{section}.\arabic{subsection}}
\renewcommand{\theequation}{S\arabic{equation}}
\renewcommand{\thefigure}{S\arabic{figure}}
\renewcommand{\thethm}{S\arabic{thm}}
\renewcommand{\theHequation}{S\arabic{equation}}
\renewcommand{\theHfigure}{S\arabic{figure}}
\renewcommand{\theHthm}{S\arabic{thm}}
\begin{center}

\newpage

{\large\bfseries Supplementary Material}\par\medskip
{\bfseries Emergent Anyons in a Weakly Interacting Gauged Chern Insulator}
\end{center}

\section{Fermionic reflection positivity: algebra and general estimates}\label{sm:general}

Here we shall extend the reflection-positivity strategy of Lieb's flux-phase theorem \cite{LiebFlux} to a setting that applies to the QWZ model. A main difference with respect to \cite{LiebFlux, GP, BGP} is the definition of {\it twisted product}, Definition \ref{def:twisted}, which allows us to simplify the proof of the reflection bound in Theorem \ref{rp}.

Fix a bond reflection $\vartheta_\nu$ of the finite lattice torus $\L_L$,
with $\nu\in\{1,2\}$, whose fixed hypersurface lies between lattice sites.
On the torus, this hypersurface has two connected components. Let
\[
 \L_L=\L_L^+\sqcup\L_L^-,\qquad
 \vartheta_\nu(\L_L^\pm)=\L_L^\mp.
\]
Let $\mathcal A$ be the full CAR algebra, and let $\mathcal A_\pm$ be its
subalgebras supported on $\L_L^\pm$. The associated Fock spaces are denoted
by $\mathcal F$ and $\mathcal F_\pm$, respectively.

The reflection $\Theta_\nu:\mathcal A\to\mathcal A$ is a
parity-preserving, antilinear $*$-automorphism satisfying
\[
 \Theta_\nu^2=\mathrm{id},\qquad
 \Theta_\nu(\mathcal A_\pm)=\mathcal A_\mp.
\]
We suppress the subscript $\nu$ when no confusion can arise.
An operator is homogeneous if it has a definite fermion parity
$p(A)\in\{0,1\}$. Operators supported on opposite halves satisfy
\[
 X_-Y_+=(-1)^{p(X_-)p(Y_+)}Y_+X_-
\]
whenever $X_-$ and $Y_+$ are homogeneous.

\begin{defn}[Twisted product]\label{def:twisted}
For homogeneous $X_-\in\mathcal A_-$ and $Y_+\in\mathcal A_+$, define
\[
 X_-\circ Y_+\doteq i^{p(X_-)p(Y_+)}X_-Y_+,
\]
and extend this operation bilinearly to
$\mathcal A_-\times\mathcal A_+$. Consequently,
\[
 \Theta(A)\circ B=i^{p(A)p(B)}\Theta(A)B
\]
for homogeneous $A,B\in\mathcal A_+$. The map
$(A,B)\mapsto\Theta(A)\circ B$ is antilinear in $A$ and linear in $B$.
\end{defn}

\begin{lemma}\label{rp-trace}
For every homogeneous $A\in\mathcal A_+$,
\[
 \Tr_{\mathcal F}(\Theta(A)\circ A)
   =\left|\Tr_{\mathcal F_+}(A)\right|^2\geq0.
\]
\end{lemma}
\begin{proof}
The graded tensor product decomposition of the finite CAR algebra gives
\[
 \Tr_{\mathcal F}(X_-Y_+)
 =\Tr_{\mathcal F_-}(X_-)\Tr_{\mathcal F_+}(Y_+).
\]
For even operators this is ordinary trace factorization. If either factor
is odd, both sides vanish. Since $\Theta$ is an antilinear $*$-isomorphism
between the two half-algebras,
\[
 \Tr_{\mathcal F_-}(\Theta(A))
   =\overline{\Tr_{\mathcal F_+}(A)}.
\]
Thus
\[
 \Tr_{\mathcal F}(\Theta(A)\circ A)
   =i^{p(A)}\left|\Tr_{\mathcal F_+}(A)\right|^2.
\]
If $A$ is even, the phase equals one. If $A$ is odd, its trace vanishes:
with $P_+\doteq(-1)^{N_+}$,
\[
 \Tr_{\mathcal F_+}(A)
 =\Tr_{\mathcal F_+}(P_+AP_+)
 =-\Tr_{\mathcal F_+}(A).
\]
This proves the claim in both cases.
\end{proof}

\Needspace{9\baselineskip}
\begin{lemma}\label{multcone}
Let $A,B,C,D\in\mathcal A_+$ be homogeneous, with
$p(A)=p(B)$ and $p(C)=p(D)$. Then
\[
 (\Theta(A)\circ B)(\Theta(C)\circ D)
   =\Theta(AC)\circ BD.
\]
\end{lemma}
\begin{proof}
Set $a=p(A)=p(B)$ and $c=p(C)=p(D)$. By the fermionic anticommutation relations:
\[
 (\Theta(A)\circ B)(\Theta(C)\circ D)
   =i^{a+c}(-1)^{ac}\Theta(AC)BD.
\]
This concludes the proof since $p(AC)=p(BD)=a+c-2ac$ and $p(AC)^2 = p(AC)$.
\end{proof}

\begin{thm}[Reflection-positive Hamiltonians]\label{rp}
Let $H$ be self-adjoint and suppose that
\begin{equation}\label{eq:rpH}
 -H=-H_+-\Theta(H_+)+\sum_\alpha\Theta(C_\alpha)\circ C_\alpha,
\end{equation}
where $H_+=H_+^*\in\mathcal A_+$ is even and the finite family
$C_\alpha\in\mathcal A_+$ consists of odd operators. Then, for every
$A\in\mathcal A_+$,
\begin{equation}\label{eq:reflectionbound}
 \Tr_{\mathcal F}\bigl(e^{-H}(\Theta(A)\circ A)\bigr)\geq0.
\end{equation}
\end{thm}
\begin{remark}
The rewriting in Eq. (\ref{eq:rpH}) should be compared with \cite{LiebFlux, GP, BGP}, where the circle product is replaced by the usual product (and the set of $C_{\alpha}$ is defined such that the Hamiltonian is Hermitian). Our representation of the Hamiltonian can be viewed as a different gauge choice with respect to \cite{LiebFlux, GP, BGP}, which simplifies the proof of the reflection bound (\ref{eq:reflectionbound}), thanks to Lemma \ref{multcone}.
\end{remark}
\begin{proof}
We denote $Q(A) = \Theta(A)\circ A$ for any $A\in\mathcal{A}_+$ and note that Lemma~\ref{multcone} reads $Q(A)Q(B)=Q(AB)$. Set $H_0=H_++\Theta(H_+)$ and
$V=\sum_\alpha Q(C_\alpha)$, so that $-H=-H_0+V$. Letting $\Delta_n=\{(t_1,\ldots,t_n):0\leq t_1\leq\cdots\leq t_n\leq1\}$, the norm-convergent Duhamel expansion is
\begin{align*}
 e^{-H}
 &=\sum_{n\geq0}\sum_{\alpha_1,\ldots,\alpha_n}
   \int_{\Delta_n}d^nt\,
   e^{-t_1H_0}Q(C_{\alpha_1})
   e^{-(t_2-t_1)H_0}\cdots
   Q(C_{\alpha_n})e^{-(1-t_n)H_0}.
\end{align*}
The $n=0$ term is understood to be $e^{-H_0}$.
The two even half-Hamiltonians commute, and hence
\[
 e^{-sH_0}=\Theta(e^{-sH_+})\circ e^{-sH_+} = Q(e^{-sH_+})
\]
Define
\[
 B_{\boldsymbol\alpha,\boldsymbol t}
 =e^{-t_1H_+}C_{\alpha_1}e^{-(t_2-t_1)H_+}
   \cdots C_{\alpha_n}e^{-(1-t_n)H_+},
\]
with $B=e^{-H_+}$ when $n=0$. Repeated application of
Lemma~\ref{multcone} gives
\[
 e^{-H}=\sum_{n\geq0}\sum_{\boldsymbol\alpha}
 \int_{\Delta_n}d^nt\,Q(B_{\boldsymbol\alpha,\boldsymbol t}).
\]
For homogeneous $A$, Lemmas~\ref{multcone} and~\ref{rp-trace} imply
\begin{align*}
 \Tr_{\mathcal F}(e^{-H}Q(A))
 &=\sum_{n\geq0}\sum_{\boldsymbol\alpha}
   \int_{\Delta_n}d^nt\,
   \left|\Tr_{\mathcal F_+}
      (B_{\boldsymbol\alpha,\boldsymbol t}A)\right|^2\geq0.
\end{align*}
For arbitrary $A=A_0+A_1$, with $p(A_j)=j$, the two mixed terms
in $Q(A)$ are odd. Since $e^{-H}$ is even, their traces vanish, and
\[
 \Tr_{\mathcal F}(e^{-H}Q(A))
 =\Tr_{\mathcal F}(e^{-H}Q(A_0))
  +\Tr_{\mathcal F}(e^{-H}Q(A_1))\geq0,
\]
concluding the proof.
\end{proof}

\begin{remark}[Self-adjointness and inverse temperature]
For odd $C$,
\[
 (\Theta(C)\circ C)^*=\Theta(C^*)\circ C^*.
\]
Thus closure of the family $\{C_\alpha\}$ under adjoints, with equal
weights for each adjoint pair, is sufficient to make the interaction
self-adjoint. This is different from \cite{LiebFlux}, where the family $\{C_{\alpha}\}$ only includes annihilation operators.
\end{remark}

\begin{thm}[Reflection bound]\label{rp-chessboard}
Let $H_+,H_-\in\mathcal A_+$ be even and self-adjoint. Suppose that
\begin{equation}\label{eq:rpform}
 -H=-H_+-\Theta(H_-)+\sum_\alpha\Theta(C_\alpha)\circ D_\alpha,
\end{equation}
where $C_\alpha,D_\alpha\in\mathcal A_+$ are odd. Define
\begin{align*}
 -H^{(-)}&=-H_--\Theta(H_-)
              +\sum_\alpha\Theta(C_\alpha)\circ C_\alpha,\\
 -H^{(+)}&=-H_+-\Theta(H_+)
              +\sum_\alpha\Theta(D_\alpha)\circ D_\alpha,
\end{align*}
and $Z=\Tr_{\mathcal F}(e^{-H})$,
$Z^{(\pm)}=\Tr_{\mathcal F}(e^{-H^{(\pm)}})$. Then
\[
 |Z|^2\leq Z^{(-)}Z^{(+)}.
\]
\end{thm}
\begin{proof}
For each $(n,\boldsymbol\alpha,\boldsymbol t)$, define
\begin{align*}
 B_-&=e^{-t_1H_-}C_{\alpha_1}e^{-(t_2-t_1)H_-}
           \cdots C_{\alpha_n}e^{-(1-t_n)H_-},\\
 B_+&=e^{-t_1H_+}D_{\alpha_1}e^{-(t_2-t_1)H_+}
           \cdots D_{\alpha_n}e^{-(1-t_n)H_+}.
\end{align*}
For $n=0$, set $B_\pm=e^{-H_\pm}$. Duhamel expansion and
Lemmas~\ref{multcone} and~\ref{rp-trace} yield
\[
 Z=\sum_{n\geq0}\sum_{\boldsymbol\alpha}\int_{\Delta_n}d^nt\,
   \overline{\Tr_{\mathcal F_+}(B_-)}
   \Tr_{\mathcal F_+}(B_+)
\]
as in the proof of Theorem~\ref{rp}. Here odd $B_\pm$ give zero, while even $B_\pm$ allow ordinary trace
factorization. By Cauchy--Schwarz inequality
\begin{equation}
 |Z|^2\leq \left(\sum_{n\geq0}\sum_{\boldsymbol\alpha}
       \int_{\Delta_n}d^nt\,
       |\Tr_{\mathcal F_+}(B_-)|^2\right)
 \left(\sum_{n\geq0}\sum_{\boldsymbol\alpha}
       \int_{\Delta_n}d^nt\,
       |\Tr_{\mathcal F_+}(B_+)|^2\right).
\end{equation}
The two factors are the Duhamel expansions of $Z^{(-)}$ and $Z^{(+)}$.
\end{proof}

\section{Application to the QWZ Hamiltonian}\label{sm:qwz}

Here we will show how to recast the Hamiltonian of the QWZ model as in Eq. (\ref{eq:rpH}). As discussed below, this will be possible after a suitable gauge fixing, different from the one of \cite{LiebFlux, GP, BGP}, and a unitary transformation.

Let $\tau^1,\tau^2,\tau^3$ be the standard Pauli matrices, and let
$c^+_{x,\alpha}=(c^-_{x,\alpha})^*$, with $\alpha\in\{1,2\}$.
The QWZ Hamiltonian without interactions is given in explicit form by
\begin{align}
 H_0(\sigma)=&m\sum_{x,\alpha,\beta}
             c^+_{x,\alpha}(\tau^3)_{\alpha\beta}c^-_{x,\beta}\nonumber-\frac t2\sum_{x,\mu,\alpha,\beta}\sigma_{x,x+e_\mu}
 \left[c^+_{x,\alpha}(D_\mu)_{\alpha\beta}c^-_{x+e_\mu,\beta}
       +c^+_{x+e_\mu,\alpha}(D_\mu^*)_{\alpha\beta}c^-_{x,\beta}\right],
 \label{qwz-rp-explicit}
\end{align}
where $D_\mu=\tau^3-i\tau^\mu$, the sums run over $x,y\in \L_L$ and 
$\alpha,\beta,\mu\in\{1,2\}$, $m\in\mathbb R$, $t\geq0$, and
$\sigma_{xy}=\sigma_{yx}\in\{\pm1\}$ is a fixed background. We also recall the definition~\eqref{eq:reflection} of the antilinear reflection automorphism
\begin{equation}\label{reflqwz}
 \Theta_\nu(c^+_{x,\alpha})
   =\sum_\beta\tau^\nu_{\alpha\beta}
                    c^-_{\vartheta_\nu(x),\beta},\qquad
 \Theta_\nu(c^-_{x,\alpha})
   =\sum_\beta c^+_{\vartheta_\nu(x),\beta}
                    \tau^\nu_{\beta\alpha},
\end{equation}
extended antilinearly and multiplicatively to the whole algebra. Hermiticity and unitarity of $\tau^\nu$ ensure that this extension preserves
adjoints and the CAR. Moreover,
\begin{equation*}
 \Theta_\nu^2(c^+_{x,\alpha})
 =\sum_{\beta,\gamma}
   \overline{\tau^\nu_{\alpha\beta}}\tau^\nu_{\gamma\beta}
    c^+_{x,\gamma}=c^+_{x,\alpha},
\end{equation*}
and similarly $\Theta_\nu^2(c^-_{x,\alpha}) = c^-_{x,\alpha}$. Thus $\Theta_\nu$ is an involutive, parity-preserving antilinear
$*$-automorphism exchanging $\mathcal A_+$ and $\mathcal A_-$.

\begin{figure}
\centering
\input{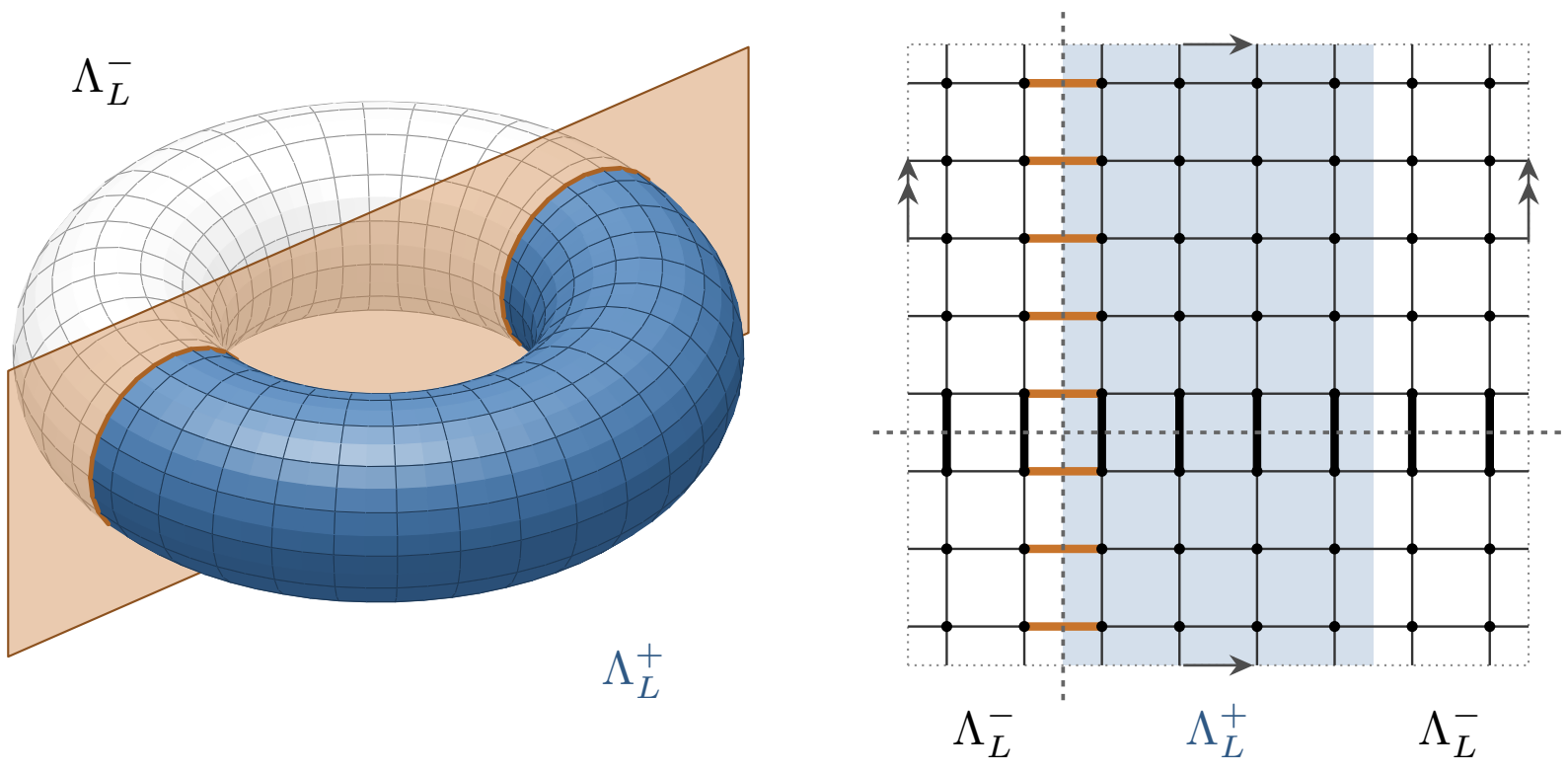}
 \caption{Reflection on the torus. The reflection plane is shaded in brown.
In the lattice diagram, thin bonds carry $\sigma=1$ and thick bonds carry
$\sigma=-1$. The thick bonds form twisting lines realizing nontrivial
$\mathbb Z_2$-holonomies, or equivalently antiperiodic boundary conditions,
in the two coordinate directions. For the vertical reflection, the shaded
strip represents $\L_L^+$ and the vertical twisting line lies along its left
boundary. A twisting line can be moved by a unitary gauge transformation.}
 \label{fig:cut}
\end{figure}

\Needspace{13\baselineskip}
\begin{proposition}\label{rp-qwz}
Fix $\nu\in\{1,2\}$. Assume that the background field is reflection
invariant on unoriented bonds:
\[
 \sigma_{\vartheta_\nu(x),\vartheta_\nu(y)}=\sigma_{xy}.
\]
Assume that $\sigma=1$ on all edges across the right boundary of $\L_L^+$, while $\sigma=-1$ across its left boundary, see Figure~\ref{fig:cut}. Then $H_0(\sigma)$ has a decomposition as required by Theorem~\ref{rp}
with reflection $\Theta_\nu$.
\end{proposition}

\begin{proof}
Write $H_0(\sigma) =H_++H_-^{\mathrm{phys}}+H_\partial$, where
\begin{align*}
 H_+={}&m\sum_{\substack{x\in\L_L^+\\\alpha,\beta}}
     c^+_{x,\alpha}(\tau^3)_{\alpha\beta}c^-_{x,\beta} -\frac t2
 \sum_{\substack{x\in\L_L^+,\ \mu=1,2\\x+e_\mu\in\L_L^+}}
 \sum_{\alpha,\beta}\sigma_{x,x+e_\mu}
 \left[c^+_{x,\alpha}(D_\mu)_{\alpha\beta}c^-_{x+e_\mu,\beta}
       +c^+_{x+e_\mu,\alpha}(D_\mu^*)_{\alpha\beta}c^-_{x,\beta}\right],
\end{align*}
and $H_-^{\mathrm{phys}}$ is the analogous restriction to $\L_L^-$.
The notation $H_-^{\mathrm{phys}}$ distinguishes this actual
negative-half operator from $H_-$ used in
Theorem~\ref{rp-chessboard}. For any $2\times2$ matrix $D$, a direct CAR calculation gives
\begin{equation}\label{rp-bilinear-reflection}
 \Theta_\nu\!\left(\sum_{\alpha,\beta}
       c^+_{x,\alpha}D_{\alpha\beta}c^-_{y,\beta}\right)
 =\delta_{xy}\,\overline{\operatorname{tr}D}
 -\sum_{\alpha,\beta}c^+_{\vartheta_\nu(y),\alpha}
       (\tau^\nu D^*\tau^\nu)_{\alpha\beta}
       c^-_{\vartheta_\nu(x),\beta}.
\end{equation}
Here $D^*$ denotes the conjugate transpose. With $D=\tau^3$, the
trace term vanishes and $-\tau^\nu\tau^3\tau^\nu=\tau^3$,
so the mass term is mapped to the mass term on the other half. For the hopping terms, set $s_{\mu\nu}=(-1)^{\delta_{\mu\nu}}$. Then
\[
 \vartheta_\nu(x+e_\mu)
   =\vartheta_\nu(x)+s_{\mu\nu}e_\mu,\qquad
 -\tau^\nu(\tau^3+i\tau^\mu)\tau^\nu
   =\tau^3+i s_{\mu\nu}\tau^\mu.
\]
Consequently, for $y=x+e_\mu$,
\[
 \Theta_\nu\!\left(\sum_{\alpha,\beta}
       c^+_{x,\alpha}(D_\mu)_{\alpha\beta}c^-_{y,\beta}\right)
 =\sum_{\alpha,\beta}c^+_{\vartheta_\nu(y),\alpha}
       (\tau^3+i s_{\mu\nu}\tau^\mu)_{\alpha\beta}
       c^-_{\vartheta_\nu(x),\beta}.
\]
For $\mu=\nu$, this is a forward hopping term on the reflected bond;
for $\mu\ne\nu$, it is the adjoint hopping term. Reflection invariance
of $\sigma$ therefore gives
\[
 \Theta_\nu(H_+)=H_-^{\mathrm{phys}}.
\]
It remains to cast the boundary terms in the required form. Set
\[
 \mathcal R_\nu=\{z\in\L_L^+:z+e_\nu\in\L_L^-\},\qquad
 \mathcal L_\nu=\{z\in\L_L^+:z-e_\nu\in\L_L^-\}.
\]
For these boundary sites,
$\vartheta_\nu(z)=z+e_\nu$ on $\mathcal R_\nu$ and
$\vartheta_\nu(z)=z-e_\nu$ on $\mathcal L_\nu$.
The inverse generator relations are
\[
 c^-_{\vartheta_\nu(z),\alpha}
 =\sum_\beta\tau^\nu_{\alpha\beta}\Theta_\nu(c^+_{z,\beta}),\qquad
 c^+_{\vartheta_\nu(z),\alpha}
 =\sum_\beta\tau^\nu_{\beta\alpha}\Theta_\nu(c^-_{z,\beta}).
\]
Let
\[
 S_\nu = -i\tau^\nu\tau^3
       =\sum_{r=1}^3\epsilon_{\nu3r}\tau^r,\qquad
 A_\nu^R=\mathbbm{1}+S_\nu,\quad A_\nu^L=\mathbbm{1}-S_\nu,
 \quad B_\nu^b=(A_\nu^b)^T\quad(b=R,L).
\]
Both $S_\nu$ and $A_\nu^b$ are Hermitian; hence
$B_\nu^b=\overline{A_\nu^b}$. For $z\in\mathcal R_\nu$, the contribution to $-H_\partial$ is
\begin{align*}
 &\frac t2\sum_{\alpha,\beta}
 \left[c^+_{z,\alpha}(D_\nu)_{\alpha\beta}c^-_{\vartheta_\nu(z),\beta}
     +c^+_{\vartheta_\nu(z),\alpha}(D_\nu^*)_{\alpha\beta}c^-_{z,\beta}\right]\\
 &\quad=\frac t2\sum_{\alpha,\beta}
 \left[(A_\nu^R)_{\alpha\beta}
          \Theta_\nu(c^-_{z,\alpha})\circ c^-_{z,\beta}
       +(B_\nu^R)_{\alpha\beta}
          \Theta_\nu(c^+_{z,\alpha})\circ c^+_{z,\beta}\right].
\end{align*}
For $z\in\mathcal L_\nu$, the oriented bond starts at
$\vartheta_\nu(z)$ and carries $\sigma=-1$. Its contribution is
\begin{align*}
 &-\frac t2\sum_{\alpha,\beta}
 \left[c^+_{\vartheta_\nu(z),\alpha}(D_\nu)_{\alpha\beta}c^-_{z,\beta}
     +c^+_{z,\alpha}(D_\nu^*)_{\alpha\beta}c^-_{\vartheta_\nu(z),\beta}\right]\\
 &\quad=\frac t2\sum_{\alpha,\beta}
 \left[(A_\nu^L)_{\alpha\beta}
          \Theta_\nu(c^-_{z,\alpha})\circ c^-_{z,\beta}
       +(B_\nu^L)_{\alpha\beta}
          \Theta_\nu(c^+_{z,\alpha})\circ c^+_{z,\beta}\right].
\end{align*}
To check the phases explicitly, the annihilation matrices are
\[
 A_\nu^R=-i\tau^\nu(\tau^3+i\tau^\nu)=\mathbbm{1}+S_\nu,
 \qquad
 A_\nu^L=i\tau^\nu(\tau^3-i\tau^\nu)=\mathbbm{1}-S_\nu,
\]
and the corresponding creation matrices are their transposes. In particular,
\begin{equation*}
A_1^R=\begin{pmatrix}1&i\\-i&1\end{pmatrix},\quad
 A_1^L=\begin{pmatrix}1&-i\\i&1\end{pmatrix},\quad
 A_2^R=\begin{pmatrix}1&1\\1&1\end{pmatrix},\quad
 A_2^L=\begin{pmatrix}1&-1\\-1&1\end{pmatrix}.
\end{equation*}
Since $S_\nu^2=\mathbbm{1}$ and $\operatorname{tr}S_\nu=0$, all four
$A$ matrices and their transposes are positive semidefinite, with
spectrum $\{0,2\}$.

For each boundary $b\in\{R,L\}$ choose a unitary $U_\nu^b$ satisfying
\[
 U_\nu^b A_\nu^b(U_\nu^b)^\dagger
   =\operatorname{diag}(0,2),
\]
and define one consistent CAR basis by
\[
 d_{z,a}^{-,b}=\sum_\alpha(U_\nu^b)_{a\alpha}c^-_{z,\alpha},
 \qquad
 d_{z,a}^{+,b}=\sum_\alpha\overline{(U_\nu^b)_{a\alpha}}
                         c^+_{z,\alpha}
                  =(d_{z,a}^{-,b})^*.
\]
Unitarity preserves the CAR. Antilinearity of $\Theta_\nu$ gives
\begin{align*}
 \sum_{\alpha,\beta}(A_\nu^b)_{\alpha\beta}
    \Theta_\nu(c^-_{z,\alpha})\circ c^-_{z,\beta}
 &=\sum_a\lambda_a\Theta_\nu(d_{z,a}^{-,b})\circ d_{z,a}^{-,b},\\
 \sum_{\alpha,\beta}(B_\nu^b)_{\alpha\beta}
    \Theta_\nu(c^+_{z,\alpha})\circ c^+_{z,\beta}
 &=\sum_a\lambda_a\Theta_\nu(d_{z,a}^{+,b})\circ d_{z,a}^{+,b},
\end{align*}
where $(\lambda_1,\lambda_2) = (0,2)$. For $a=1,2$, now define
\[
 C_{z,a}^{\pm,b}\doteq\sqrt{\frac{t\lambda_a}{2}}\,
                           d_{z,a}^{\pm,b}.
\]
With $\mathcal B_R=\mathcal R_\nu$ and $\mathcal B_L=\mathcal L_\nu$,
the full boundary interaction becomes
\[
 -H_\partial=
 \sum_{b=R,L}\sum_{z\in\mathcal B_b}\sum_{a=1}^2
 \left[\Theta_\nu(C_{z,a}^{-,b})\circ C_{z,a}^{-,b}
      +\Theta_\nu(C_{z,a}^{+,b})\circ C_{z,a}^{+,b}\right].
\]
Thus the interaction is in the required reflection-positive form, and Theorem~\ref{rp} applies.
\end{proof}

For an arbitrary bond background $\sigma$, let $\varepsilon_x\in\{\pm1\}$, supported on $\mathcal R_\nu\cup\mathcal L_\nu$, be such that the gauge-transformed field $\sigma^\varepsilon_{xy}=\varepsilon_x\sigma_{xy}\varepsilon_y$ equals $+1$ on the bonds leaving $\mathcal R_\nu$ and $-1$ on the bonds leaving $\mathcal L_\nu$. Then, writing again $H_+$ and $H_-^{\rm phys}$ for the restrictions of $H_0(\sigma^\varepsilon)$ to $\Lambda_L^\pm$ and $H_- = \Theta_\nu(H_-^{\rm phys})\in\mathcal A_+$, the boundary computation in the proof of Proposition~\ref{rp-qwz} applies unchanged and gives a representation of the Hamiltonian as in Eq. (\ref{eq:rpform}),
\[
  -H_0(\sigma^\varepsilon) = -H_+ - \Theta_\nu(H_-) + \sum_j \Theta_\nu(C_j)\circ C_j
\]
with the $\sigma$-independent operators $C_j$ of Proposition~\ref{rp-qwz}. Thus, Theorem~\ref{rp-chessboard} with $D_j=C_j$ and gauge invariance of the trace yield $Z_\lambda(\sigma)^2\le Z_\lambda(\sigma^-)\,Z_\lambda(\sigma^+)$, where $\sigma^\pm$ is the reflection-invariant background that agrees with $\sigma^\varepsilon$ on $\Lambda_L^\pm$.

\section{Local Hubbard interaction and reflection positivity}
\label{sm:hubbard}

The many-body interaction is
\begin{equation}
 \lambda V=\lambda\sum_x(n_{x,1}-\tfrac12)(n_{x,2}-\tfrac12)
 =\frac{\lambda}{2}\sum_x(N_x-1)^2-\frac{\lambda}{4}L^2.
 \label{sm:hubbard-def}
\end{equation}
It is number conserving and invariant under an on-site change of
orbital basis. Since $\tau^\nu$ is off diagonal for $\nu=1,2$, the CAR and~\eqref{reflqwz} gives
\begin{equation}
 \Theta_\nu(n_{x,1})=1-n_{\vartheta_\nu(x),2},\qquad
 \Theta_\nu(n_{x,2})=1-n_{\vartheta_\nu(x),1}.
\end{equation}
Consequently $\Theta_\nu((n_{x,1}-\frac{1}{2})(n_{x,2}-\frac{1}{2}))
=(n_{\vartheta_\nu(x),1}-\frac{1}{2})(n_{\vartheta_\nu(x),2}-\frac{1}{2})$, since the two density operators commute. Write $V_+=\sum_{x\in\Lambda^+}(n_{x,1} - \frac{1}{2})(n_{x,2} - \frac{1}{2})$, and let $H_+(\lambda)=H_++\lambda V_+$.
Adding $\lambda V_+$ to the positive-half Hamiltonian in
Proposition~\ref{rp-qwz} leaves its boundary couplings unchanged and yields
\begin{equation}
 -H_\lambda=-H_+(\lambda)-\Theta_\nu(H_+(\lambda))
       +\sum_j\Theta_\nu(C_j)\circ C_j,
\end{equation}
which is again in RP form.

\begin{figure}
\centering
\begin{tikzpicture}[x=0.6cm, y=0.6cm, font=\small]
  \tikzset{
    bond/.style={black!80, line width=0.45pt},
    neg/.style={black, line width=1.6pt, line cap=butt},
    idarrow/.style={-{Stealth[length=1.6mm,width=1.4mm]}, black!70, line width=0.5pt},
    idarrow2/.style={-{Stealth[length=1.6mm,width=1.4mm] Stealth[length=1.6mm,width=1.4mm]},
                     black!70, line width=0.5pt},
  }
  % \flatpanel{x-shift}{a}{b}{label}
  \newcommand{\flatpanel}[4]{%
    \begin{scope}[shift={(#1,0)}]
      \draw[black!45, densely dotted, line width=0.4pt] (0,0) rectangle (4,4);
      \foreach \k in {0,...,3} {
        \draw[bond] (0,\k+0.5) -- (4,\k+0.5);
        \draw[bond] (\k+0.5,0) -- (\k+0.5,4);
      }
      % a = -1: horizontal wrap bonds (crossing the left/right edge)
      \ifnum#2<0
        \foreach \k in {0,...,3} {
          \draw[neg] (0,\k+0.5) -- (0.5,\k+0.5);
          \draw[neg] (3.5,\k+0.5) -- (4,\k+0.5);
        }
      \fi
      % b = -1: vertical wrap bonds (crossing the bottom/top edge)
      \ifnum#3<0
        \foreach \k in {0,...,3} {
          \draw[neg] (\k+0.5,0) -- (\k+0.5,0.5);
          \draw[neg] (\k+0.5,3.5) -- (\k+0.5,4);
        }
      \fi
      \foreach \i in {0,...,3} \foreach \j in {0,...,3}
        { \fill[black] (\i+0.5,\j+0.5) circle (1.1pt); }
      \foreach \y in {0,4} { \draw[idarrow] (1.6,\y) -- (2.4,\y); }
      \foreach \x in {0,4} { \draw[idarrow2] (\x,1.55) -- (\x,2.45); }
      \node[below=3pt] at (2,0) {#4};
    \end{scope}}
  \flatpanel{0}{1}{1}{$(a,b)=(1,1)$}
  \flatpanel{5.2}{-1}{1}{$(a,b)=(-1,1)$}
  \flatpanel{10.4}{1}{-1}{$(a,b)=(1,-1)$}
  \flatpanel{15.6}{-1}{-1}{$(a,b)=(-1,-1)$}
\end{tikzpicture}
 \caption{Representatives of the four flat gauge classes
on a torus with $L=4$. Every plaquette has $B_p=+1$.
Thin and thick bonds denote $\sigma_{xy}=+1$ and
$\sigma_{xy}=-1$, respectively.
The labels $(a,b)$ give the
holonomies in the horizontal and vertical directions.}
 \label{fig:flat-sectors}
\end{figure}
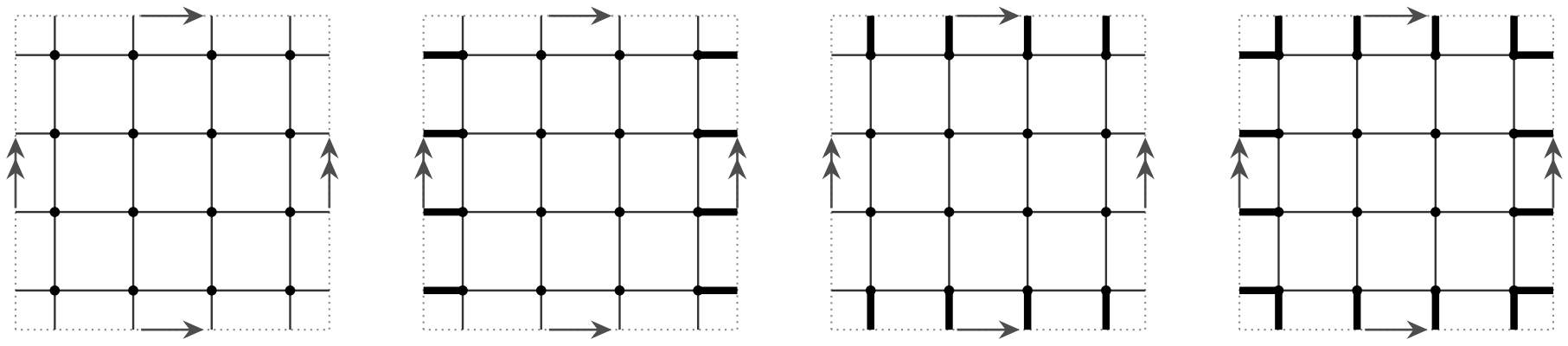

\section{Reflection positivity and chessboard estimates}
\label{sm:defects}

Let $Z_{\beta,L}(\Phi;a,b;\lambda)$ be the partition function of the fermionic Hamiltonian, where $\Phi=(B_{x,y})$ specifies the fluxes across all plaquettes and $a,b$ the two holonomies. Define
\begin{equation}
 \mathcal F(\Phi)=-\log\max_{a,b=\pm1}
                    Z_{\beta,L}(\Phi;a,b;\lambda).
 \label{sm:F-definition}
\end{equation}
On admissible configurations ($\prod_pB_p=1$) this value is finite, and we extend $\mathcal F$ by
$+\infty$ on inadmissible configurations.

\subsection{Minimizing magnetic configuration}

Without loss of generality, we assume that the spins associated to bonds intersecting the right boundary of $\Lambda_{L}^{+}$ are equal to $+1$, while spins associated to bonds intersecting the left boundary of $\Lambda_{L}^{+}$ are equal to $-1$. This configuration can be realized after a suitable gauge transformation, that does not change the partition function. Next, we apply Theorem~\ref{rp-chessboard} to these two halves.

Let us focus on the plaquettes at a seam. There, after the reflection, the two
normal bonds have identical signs, while the two tangential bonds
are reflected copies. Therefore
\begin{equation}
 B_{\rm seam}=(\pm1)^2\sigma_{\rm tangent}^2=+1.
 \label{sm:seam-zero}
\end{equation}
This is different from \cite{LiebFlux, GP, BGP}, where the plaquettes at the seam have $\pi$ flux rather than $0$ flux. Next, denote columns of plaquettes by $X^j=(B_{j,y})_{y=0}^{L-1}$ and let
$\mathbf1=(1,\ldots,1)$. By Theorem \ref{rp-chessboard}:
\begin{align}
 \mathcal F(X^1,\ldots,X^L)\geq\tfrac12\{&
 \mathcal F(X^1,\ldots,X^{L/2-1},\mathbf1,
                  X^{L/2-1},\ldots,X^1,\mathbf1)\nonumber\\
 &+\mathcal F(X^L,\ldots,X^{L/2+1},\mathbf1,
                  X^{L/2+1},\ldots,X^L,\mathbf1)\}.
 \label{sm:column-reflection}
\end{align}
After a reflection, the holonomies of the new configurations might differ from the original ones; taking the maximum over $a,b$ makes the inequality independent of these choices. Cyclic translation of columns or rows leaves $\mathcal F$ unchanged. Repeated application of this inequality implies that the $0$-flux configuration maximizes the partition function. Also, applying both reflections to a flat configuration fixes both normal holonomies
to $-1$; therefore,
\begin{equation}
 \max_{a,b}Z_{\beta,L}(\mathbf1;a,b;\lambda)
        =Z_{\beta,L}(\mathbf1;-1,-1;\lambda)=Z_{\rm ref}.
 \label{sm:flat-max}
\end{equation}

\subsection{Chessboard estimate}

The chessboard lemma of \cite{GP},
Sec.~3.3, Lemma~3.10, applies to \eqref{sm:column-reflection}, with
the constant seam column $-\mathbf1$ of \cite{GP} replaced by
$+\mathbf1$. We have, following \cite{GP}:
\begin{equation}
 \mathcal F(\Phi)\geq\frac1{L^2}
  \sum_{0\leq j,y<L}
       \mathcal F\bigl(\mathcal D(B_{j,y})\bigr).
 \label{sm:row-dissemination}
\end{equation}
Here $\mathcal D(q)$ places $q$ at every even-even plaquette and
$+1$ at every other plaquette. For $q=+1$ it is the flat pattern;
for $q=-1$ it is a chessboard pattern, as in Figure \ref{fig:cut1}. Every original plaquette occurs once in the sum; hence:
\begin{equation}
 \mathcal F(\Phi)\geq\frac{L^2-n_v}{L^2}\mathcal F(\mathbf1)
               +\frac{n_v}{L^2}\mathcal F(\Phi^*).
 \label{sm:CB-input}
\end{equation}
Dividing by $\beta$ and taking the limit $\beta\to \infty$, Eq. (\ref{eq:chess0}) follows, with $d_{*,\lambda} = \lim_{\beta \to \infty} d_{*,\beta,\lambda}$ and
\begin{equation}
 d_{*,\beta,\lambda} = -\frac1{\beta L^2}
   \log\frac{\max_{a,b}Z_{\beta,L}(\Phi^*;a,b;\lambda)}
                  {Z_{\rm ref}}.
 \label{sm:mass-density}
\end{equation}

\subsection{The fermionic Hamiltonian in a chessboard flux pattern}

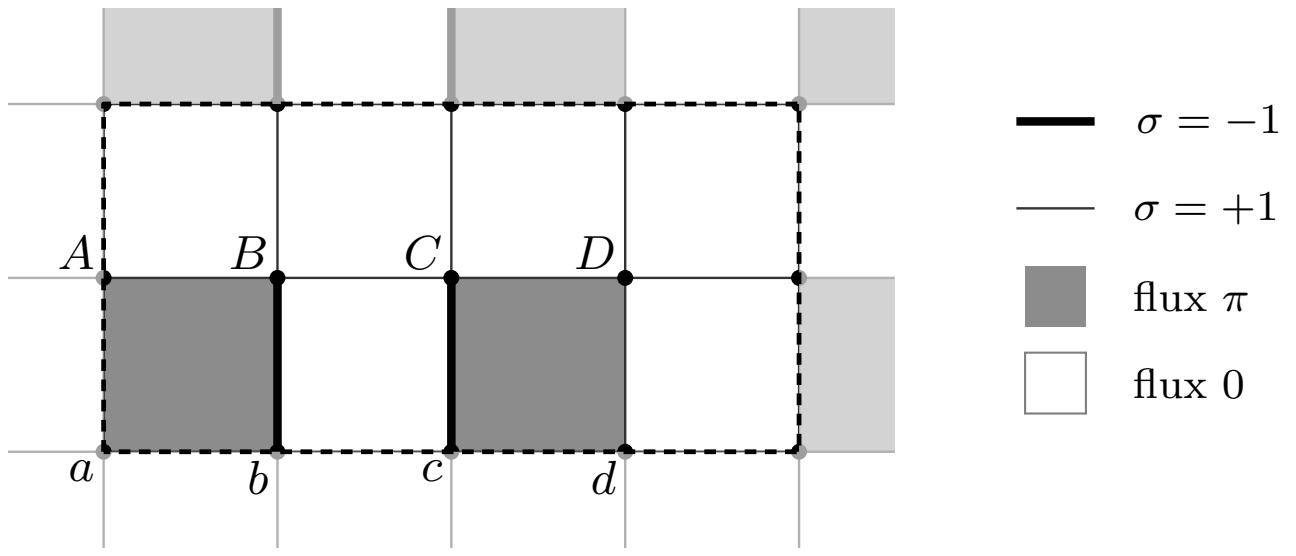
\begin{figure}
\centering
\begin{tikzpicture}[x=1.15cm, y=1.15cm, font=\small]
  \colorlet{piflux}{black!45}
  \tikzset{
    bond/.style={black!80, line width=0.45pt},
    neg/.style={black, line width=1.6pt, line cap=butt},
  }
  \def\xa{-0.55} \def\xb{4.55} \def\ya{-0.55} \def\yb{2.55}
  \begin{scope}
    \clip (\xa,\ya) rectangle (\xb,\yb);
    % pi-flux plaquettes: lower-left corner (x,y) with x, y even
    \foreach \x in {-2,0,2,4} \foreach \y in {-2,0,2}
      { \fill[piflux] (\x,\y) rectangle (\x+1,\y+1); }
    % bonds: horizontal all +1; vertical sigma_2(x,0) = (1,-1,-1,1), sigma_2(x,1) = 1
    \foreach \y in {-1,...,3} { \draw[bond] (\xa,\y) -- (\xb,\y); }
    \foreach \x in {-1,...,5} \foreach \y in {-1,...,2} {
      \pgfmathtruncatemacro{\xm}{mod(\x+8,4)}
      \pgfmathtruncatemacro{\ym}{mod(\y+8,2)}
      \ifnum\ym=0
        \ifnum\xm=1 \draw[neg] (\x,\y) -- (\x,\y+1);
        \else\ifnum\xm=2 \draw[neg] (\x,\y) -- (\x,\y+1);
        \else \draw[bond] (\x,\y) -- (\x,\y+1); \fi\fi
      \else \draw[bond] (\x,\y) -- (\x,\y+1); \fi
    }
    \foreach \x in {-1,...,5} \foreach \y in {-1,...,3}
      { \fill[black] (\x,\y) circle (1.5pt); }
  \end{scope}
  % fade everything outside the cell [0,4) x [0,2)
  \fill[white, fill opacity=0.62, even odd rule]
    (\xa-0.05,\ya-0.05) rectangle (\xb+0.05,\yb+0.05) (0,0) rectangle (4,2);
  % the cell
  \draw[black, line width=0.9pt, dash pattern=on 2.2pt off 1.6pt, line join=round]
    (0,0) rectangle (4,2);
  % site labels
  \foreach \x/\l in {0/a,1/b,2/c,3/d}
    { \node[anchor=north east, inner sep=1.5pt] at (\x,0) {$\l$}; }
  \foreach \x/\l in {0/A,1/B,2/C,3/D}
    { \node[anchor=south east, inner sep=1.5pt] at (\x,1) {$\l$}; }
  % legend
  \begin{scope}[shift={(5.25,0.05)}, font=\footnotesize]
    \draw[neg] (0,1.85) -- (0.45,1.85) node[right=4pt, black] {$\sigma=-1$};
    \draw[bond] (0,1.35) -- (0.45,1.35) node[right=4pt, black] {$\sigma=+1$};
    \fill[piflux] (0.05,0.67) rectangle (0.40,1.02);
    \node[right=4pt] at (0.45,0.845) {flux $\pi$};
    \draw[black!50, line width=0.4pt] (0.05,0.17) rectangle (0.40,0.52);
    \node[right=4pt] at (0.45,0.345) {flux $0$};
  \end{scope}
\end{tikzpicture}
 \caption{Bond gauge for the disseminated defect pattern and the sites
of its magnetic cell. The boundary is periodically identified after
translations by $4e_1$ and $2e_2$. Uppercase and lowercase letters label
sites, not independent physical species.}
 \label{sm:fig-cell}
\end{figure}

Here we shall determine the Hamiltonian of the fermionic model associated with the chessboard flux configuration $\Phi^{*}$, Fig. \ref{fig:cut1}. Let us introduce a periodic $4\times2$ bond cell, with sites
$a,b,c,d$ on $y=0$ and $A,B,C,D$ on $y=1$, in increasing order of
$x=0,1,2,3$. See Figure~\ref{sm:fig-cell}. All horizontal signs are positive; vertically set
\begin{equation}
 \sigma_2(x,0)=s_x=(1,-1,-1,1)_x,\qquad \sigma_2(x,1)=1,
 \label{sm:cell-gauge}
\end{equation}
yields the fluxes depicted in Figure~\ref{sm:fig-cell}.

Put $T_\mu=-tD_\mu/2$ and $Q=m\tau^3$. Explicitly,
\begin{equation}
 D_1=\begin{pmatrix}1&-i\\-i&-1\end{pmatrix},\qquad
 D_2=\begin{pmatrix}1&-1\\1&-1\end{pmatrix},\qquad
 T_\mu^2=0,\quad\|T_\mu\|=t.
\end{equation}
For cell positions $R\in4\mathbb Ze_1+2\mathbb Ze_2$, the free
Hamiltonian is, with every bilinear summed over $\alpha,\beta$,
\begin{align}
 H_*(0)={}&\sum_{R,x,y,\alpha,\beta}
       c^+_{R+(x,y),\alpha}Q_{\alpha\beta}c^-_{R+(x,y),\beta}\nonumber\\
 &+\sum_{R,y,\alpha,\beta}\left[
  \sum_{x=0}^2c^+_{R+(x,y),\alpha}(T_1)_{\alpha\beta}
                              c^-_{R+(x+1,y),\beta}
 +c^+_{R+(3,y),\alpha}(T_1)_{\alpha\beta}c^-_{R+4e_1+(0,y),\beta}
   +\mathrm{h.c.}\right]\nonumber\\
 &+\sum_{R,x,\alpha,\beta}\left[
  s_xc^+_{R+(x,0),\alpha}(T_2)_{\alpha\beta}c^-_{R+(x,1),\beta}
 +c^+_{R+(x,1),\alpha}(T_2)_{\alpha\beta}c^-_{R+2e_2+(x,0),\beta}
   +\mathrm{h.c.}\right].
 \label{sm:cell-realspace}
\end{align}
Here $0\leq x<4$, $0\leq y<2$ whenever these indices are summed. The interacting Hamiltonian is $H_*(\lambda)=H_*(0)+\lambda V$, with $V$ the Hubbard interaction.

Let $z_j=e^{i\kappa_j}$, $\kappa_j\in[0,2\pi)$ denote cell translation
phases and set $B_x=s_xT_2+z_2^{-1}T_2^*$.
In the basis $(a,b,c,d,A,B,C,D)$, the requested Bloch matrix is
\begin{equation}
 h_*(\kappa)=\begin{pmatrix}
 Q&T_1&0&z_1^{-1}T_1^*&B_0&0&0&0\\
 T_1^*&Q&T_1&0&0&B_1&0&0\\
 0&T_1^*&Q&T_1&0&0&B_2&0\\
 z_1T_1&0&T_1^*&Q&0&0&0&B_3\\
 B_0^*&0&0&0&Q&T_1&0&z_1^{-1}T_1^*\\
 0&B_1^*&0&0&T_1^*&Q&T_1&0\\
 0&0&B_2^*&0&0&T_1^*&Q&T_1\\
 0&0&0&B_3^*&z_1T_1&0&T_1^*&Q
 \end{pmatrix}.
 \label{sm:cell-bloch}
\end{equation}
Every entry is a $2\times2$ matrix; in particular
$(h_*)_{r\alpha,r'\beta}$ retains the indices $\alpha,\beta$.
The terms $z_2^{-1}T_2^*$ are the reverse hoppings from the neighbouring
cell. Two useful checks are
\begin{equation}
 \operatorname{tr}h_*=0,\qquad
 \operatorname{tr}h_*^2=16(m^2+2t^2).
 \label{sm:cell-moments}
\end{equation}
The second follows by summing the squared hopping norms; interference
terms vanish because $T_\mu^2=0$.

\subsection{Evaluation of the fermionic trace and zero-temperature limit}

Diagonalize $h_*(\kappa)$ as $U_\kappa\operatorname{diag}
(\varepsilon_1,\ldots,\varepsilon_{16})U_\kappa^*$. For $\lambda = 0$, we get:
\begin{align}
 Z_{*,\beta,L}(0)&=
 \prod_{\kappa}\prod_{a=1}^{16}(1+e^{-\beta\varepsilon_a(\kappa)})
 =\prod_\kappa\det(1+e^{-\beta h_*(\kappa)})\nonumber\\
&=\prod_{\kappa,a}2\cosh\!\left(\frac{\beta\varepsilon_a(\kappa)}2\right),
 \label{sm:trace-exact}\\
\end{align}
The exponential prefactors disappear because $\sum_a\varepsilon_a=0$.
There are $L^2/8$ cell momenta. For holonomies $(e^{i\phi_1},e^{i\phi_2})$,
\begin{equation}
 \kappa_1=\frac{2\pi n_1+\phi_1}{L/4},\quad
 \kappa_2=\frac{2\pi n_2+\phi_2}{L/2},\qquad
 0\leq n_1<L/4,\quad0\leq n_2<L/2.
\end{equation}
For $a,b=\pm1$, $\phi_j$ is $0$ or $\pi$. The partition function of the flat sector is:
\begin{equation}
 Z_{+,\beta,L}(0)=\prod_k4\cosh^2(\beta E(k)/2).
\end{equation}

The ground state energies of the chessboard background and of the flat background are, in the infinite volume limit:
\begin{align}
 e_*(m,t)&=-\frac1{16}\int_{[0,2\pi]^2}
              \frac{\dd^2\kappa}{(2\pi)^2}\operatorname{tr}|h_*(\kappa)|,\\
 e_+(m,t)&=-\int_{[0,2\pi]^2}\frac{\dd^2k}{(2\pi)^2}E(k),\qquad
 d_*(m,t)=e_*(m,t)-e_+(m,t).
 \label{sm:margin-integral}
\end{align}
These expressions are the starting point for the numerical evaluation of the magnetic gap~(\ref{eq:mass-stability}).

\section{Quasi-local control of the ground states}
\label{sm:physical}

\subsection{Almost-insensitivity of the ground state energy on magnetic twists}

Let $E_L(\varphi)$ be the non-degenerate flat-sector fermionic ground state energy, separated from the rest of the spectrum by a uniform gap. Let $I_C(\varphi)=\partial_jH_C(\varphi)$ be the current associated with a twist $\varphi_j$ on a cut $C$. The twist can be translated to a parallel cut $C'$ at distance $L/2$ by a
site-dependent number rotation $R = R(\varphi_j)$:
\begin{equation}
 R H_C(\varphi)R^*=H_{C'}(\varphi),\qquad
 R I_C(\varphi)R^*=J_C,
\end{equation}
where $J_C$ is the same bond current with its local twist phase removed,
and hence is independent of $\varphi_j$. The on-site interaction is
unchanged. Hellmann--Feynman and unitary covariance imply
\begin{equation}
 \partial_jE_L(\varphi)=\langle J_C\rangle_{C',\varphi}.
 \label{sm:current-remote}
\end{equation}
The quasi-adiabatic generator for varying the twist of $H_{C'}$, see Sec. \ref{sec:qaflow} below, is
localized near $C'$. In particular, for every $N$:
\begin{equation}
 \left|\partial_j\langle J_C\rangle_{C',\varphi}\right|
 =\left|\langle[J_C,K_{C'}]\rangle_{C',\varphi}\right|
 \leq C_NL^2(L/2)^{-N}=O(L^{-\infty}),
 \label{sm:current-variation}
\end{equation}
where the constants are uniform in $\varphi$. Since $E_L$ is $2\pi$-periodic in $\varphi_j$,
$\int_0^{2\pi}\partial_jE_L d\varphi_j=0$.
A periodic function with vanishing average and oscillation
$O(L^{-\infty})$ has supremum $O(L^{-\infty})$. Thus
\begin{equation}
 \sup_\varphi|\partial_jE_L(\varphi)|=O(L^{-\infty}),\qquad
 \epsilon_L = \sup_{\varphi,\varphi'}|E_L(\varphi)-E_L(\varphi')|
                =O(L^{-\infty}),
 \label{sm:energy-splitting}
\end{equation}
as claimed in~\eqref{eq:GS splitting}.

\subsection{Local indistinguishability}

For a bounded local physical observable $A$ supported in a fixed-size region, choose representatives of the flat magnetic backgrounds whose nontrivial seams are at distance of order $L$ from that region. They agree on the support of the observable. By the quasi-adiabatic flow, we can connect the matter ground states by continuous twists on these seams. With $P(\varphi)$ the ground-state projector,
\begin{equation}
 \partial_j\Tr(P(\varphi)A)=i\Tr(P(\varphi)[A,K_j(\varphi)]),
 \qquad
 |\partial_j\Tr(PA)|\leq C_{A,N}L^{-N}
\end{equation}
for every $N$. Matrix elements of components that change plaquette flux vanish between flat states. Furthermore, off-diagonal matrix elements
between distinct holonomies vanish exactly. Consequently
\begin{equation}
 \Pi_L A\Pi_L=a_L(A)\Pi_L+O_A(L^{-\infty}),
 \label{sm:local-indistinguishability}
\end{equation}
where $a_{L}(A)$ is the average of $A$ over any of the four matter ground states.

\section{Quasi-adiabatic flow, the ground-state bundle, and the loop algebra}
\label{sm:braiding}

\subsection{The quasi-adiabatic flow}\label{sec:qaflow}

Let $H(\varphi)=H(\lambda;\varphi)$, with twists on two oriented dual
cuts, and let $P(\varphi)$ be the projector onto the gapped ground state in a flat magnetic background. Choose a real odd integrable function $f_g$ with superpolynomial decay,
whose Fourier transform is such that
\begin{equation}
 \widehat f_g(\omega)=\int_{\mathbb R}e^{-i\omega u}f_g(u)\dd u,
 \qquad
 \widehat f_g(\omega)=-\frac{i}{\omega}\quad(|\omega|\geq g_f),
 \qquad \widehat f_g(\omega)=0\quad(|\omega|\leq g_f/2).
 \label{sm:filter}
\end{equation}
For example take $\widehat f_g=-i(1-\chi_g)/\omega$, where
$\chi_g$ is an even smooth cutoff equal to 1 near zero and zero
outside the gap interval. The inverse is understood as an oscillatory
integral at high frequency; it is an integrable real odd function.
Its bounded jump at $u=0$ causes no difficulty, and all time-tail
moments are finite. This is the spectral-flow filter
of Refs.~\cite{HW,BMNS}. Define
\begin{equation}
 K_j(\varphi)=\int_{\mathbb R} f_g(u)
       e^{iuH(\varphi)}\partial_jH(\varphi)e^{-iuH(\varphi)}\dd u.
 \label{sm:K-explicit}
\end{equation}
This operator is the generator of the quasi-adiabatic flow. By Lieb-Robinson bounds, it follows that it is a quasi-local operator. It generates the parallel transport of the many-body matter ground state on the flux torus. We have \cite{HW,BMNS}:
\begin{equation}
 \partial_jP=i[K_j,P],\qquad PK_jP=0.
 \label{sm:flow-identity}
\end{equation}
For a piecewise smooth path $\gamma:[0,1]\to\mathbb T^2_\varphi$,
set $K_\gamma(s)=\dot\gamma_j(s)K_j(\gamma(s))$ and consider the solution of
\begin{equation}
 \dot V_\gamma(s)=iK_\gamma(s)V_\gamma(s),\qquad V_\gamma(0)=I.
 \label{sm:transport-equation}
\end{equation}
Differentiating $V_\gamma^*P(\gamma(s))V_\gamma$ with
\eqref{sm:flow-identity} gives zero, hence
\begin{equation}
 V_\gamma(s)P(\gamma(0))V_\gamma(s)^*=P(\gamma(s)).
 \label{sm:projector-intertwining}
\end{equation}

\subsection{Parallel transport in a line bundle of boundary conditions}

The ground-state fibres
\begin{equation}
 \mathcal E_\varphi=\operatorname{Ran}P (\varphi),\qquad
 \mathcal E=\bigsqcup_{\varphi\in\mathbb T^2_\varphi}\mathcal E_\varphi
 \label{sm:ground-bundle}
\end{equation}
form a complex line bundle. $H$ and $P$ are smooth and periodic in
cut twists, but a normalized periodic ground vector need not exist
globally. Choose local frames $\psi_a$ on contractible patches $U_a$.
On overlaps $\psi_b=e^{i\chi_{ab}}\psi_a$, so
\begin{equation}
 \mathcal A_a=i\langle\psi_a|\dd\psi_a\rangle,\qquad
 \mathcal A_b=\mathcal A_a-\dd\chi_{ab},\qquad
 F=\dd\mathcal A_a=i\Tr P[\partial_1P,\partial_2P]\dd^2\varphi.
\end{equation}
The curvature is globally defined and its integral is $2\pi C$.
A nonzero $C$ obstructs a global smooth periodic frame, see Fig.~\ref{fig:bundle}. On a patch write $V_\gamma(s)\psi(\gamma(0))=c(s)\psi(\gamma(s))$.
Since $PK_\gamma P=0$,
\begin{equation}
 \frac{\dd}{\dd s}\log c(s)
 =-\langle\psi(\gamma(s))|\tfrac{\dd}{\dd s}\psi(\gamma(s))\rangle.
 \label{sm:horizontal-lift}
\end{equation}
Thus the transported vector satisfies the horizontal condition
$\langle V\psi(0)|\partial_s(V\psi(0))\rangle=0$.
For a loop bounding a contractible square, Stokes' theorem gives
\begin{equation}
 V_{\partial\Gamma}P(0)=
 \exp\!\left(-\oint_{\partial\Gamma}\langle\psi|\dd\psi\rangle\right)P(0)
 =\exp\!\left(i\int_\Gamma F_L\dd^2\varphi\right)P(0).
 \label{sm:berry-phase}
\end{equation}

\subsection{Switching the ground states of the gauge theory}

Choose oriented cuts $C_j^*$ and set
\begin{equation}
 \widehat X_j=\prod_{b\pitchfork C_j^*}\widehat\sigma_b^x,\qquad
 \widehat X_jH(\varphi)\widehat X_j=H(\varphi+\pi e_j).
 \label{sm:gauge-covariance}
\end{equation}
Conjugation by $\widehat X_j$ changes the sign of each hopping across the cut, which is
equivalent to increasing its external twist by $\pi$. Plaquette fluxes
remain unchanged, because a closed dual cut crosses every plaquette boundary an even number of times. Obviously, the two $\widehat X_j$ commute. The generator of the quasi-adiabatic flow (\ref{sm:K-explicit}) satisfies: 
\begin{equation}
 \widehat X_jK_\ell(\varphi)\widehat X_j
           =K_\ell(\varphi+\pi e_j),\qquad
 \widehat X_jV_\gamma=V_{\gamma+\pi e_j}\widehat X_j.
 \label{sm:covariance}
\end{equation}
Let $\alpha_1(s)=(\pi s,0)$ and $\alpha_2(s)=(0,\pi s)$ and define
$W_j=\widehat X_jV_{\alpha_j}$. These operators intertwine the four ground states of the gauge theory:
\begin{align}
 W_1|\Omega_{a,b}\rangle&=c_1(a,b)|\Omega_{-a,b}\rangle,\\
 W_2|\Omega_{a,b}\rangle&=c_2(a,b)|\Omega_{a,-b}\rangle,
 \qquad |c_j(a,b)|=1.
 \label{sm:sector-switching}
\end{align}

\subsection{Nontrivial braiding of magnetic excitations}

Using \eqref{sm:covariance} we obtain:
\begin{align}
 W_2^*W_1^*W_2W_1
 &=V_{\alpha_2}^*V_{\alpha_1+\pi e_2}^*
                      V_{\alpha_2+\pi e_1}V_{\alpha_1}\nonumber\\
 &=V_{\partial\Gamma},\qquad\Gamma=[0,\pi]^2.
 \label{sm:commutator-path}
\end{align}
The path is $(0,0)\to(\pi,0)\to(\pi,\pi)\to(0,\pi)\to(0,0)$,
positively oriented for $\dd\varphi_1\wedge\dd\varphi_2$.
In sector $(a,b)$ the curvature integral is on
$\Gamma+\varphi_{a,b}$, with $\varphi_{a,b,j}=0$ or $\pi$.
This is the geometrical construction of \cite{BGP}, Sec. 6.1.3, this time with nonzero curvature.

Under the uniform gap assumption, the curvature $F_{L}(\varphi)$ turns out to be almost constant on the flux torus \cite{BBDF}, in the sense of Eq. (\ref{sm:flat-curvature}). Since each translated square has area $\pi^2$,
\begin{equation}
 \int_{\Gamma+\varphi_{a,b}}F_L\dd^2\varphi
     =\pi^2\frac{C}{2\pi}+O(L^{-\infty})
     =\frac{\pi C}{2}+O(L^{-\infty}).
\end{equation}
Combining this with \eqref{sm:berry-phase} proves
\begin{equation}
 \Pi_LW_2^*W_1^*W_2W_1\Pi_L
       =e^{i \frac{\pi}{2}C}\Pi_L+O(L^{-\infty}),\qquad
 W_2W_1\Pi_L=e^{i \frac{\pi}{2}C}W_1W_2\Pi_L+O(L^{-\infty}).
\end{equation}
Finally, for a Wilson loop $Z_C=\prod_{b\in C}\widehat\sigma_b^z$, proceeding as in \cite{BGP} we also have:
\begin{equation}
 Z_CW_{C^*}=(-1)^{I(C,C^*)}W_{C^*}Z_C.
\end{equation}
\end{document}